\documentclass[12pt]{iopart}

\usepackage[T1]{fontenc}

\usepackage[latin1]{inputenc}

\usepackage{graphicx}

\usepackage{multirow}

\usepackage[english]{babel}

\usepackage[usenames,dvipsnames,svgnames]{xcolor}

\usepackage[T1]{fontenc}

\usepackage{amssymb}

\usepackage{amsfonts}

\usepackage{subfig}

\usepackage{dcolumn}
\usepackage{epsf}
\usepackage{enumerate}
\usepackage{hhline}
\usepackage{array}
\usepackage{tabularx}
\usepackage{appendix}
\usepackage{hyperref}
\usepackage{epstopdf}
\usepackage{cleveref}

\Crefname{figure}{Fig.}{Figs.}

\usepackage{siunitx}

\DeclareSIUnit\year{yr}

\usepackage{titlesec}

\expandafter\let\csname equation*\endcsname\relax

\expandafter\let\csname endequation*\endcsname\relax

\usepackage{amsmath}

\newcommand{\be}{\begin{equation}}

\newcommand{\ee}{\end{equation}}

\newcommand{\bea}{\begin{eqnarray}}

\newcommand{\eea}{\end{eqnarray}}

\newcommand{\beaa}{\begin{eqnarray*}}

\newcommand{\eeaa}{\end{eqnarray*}}

\newcommand{\nn}{\nonumber \\}

\newcommand\tab[1][1cm]{\hspace*{#1}}

\titleformat{\paragraph}
{\normalfont\normalsize\bfseries}{\theparagraph}{1em}{}
\titlespacing*{\paragraph}
{0pt}{3.25ex plus 1ex minus .2ex}{1.5ex plus .2ex}

\titleclass{\subsubsubsection}{straight}[\subsection]

\newcounter{subsubsubsection}[subsubsection]
\renewcommand\thesubsubsubsection{\thesubsubsection.\arabic{subsubsubsection}}
\renewcommand\theparagraph{\thesubsubsubsection.\arabic{paragraph}} 

\titleformat{\subsubsubsection}
  {\normalfont\normalsize\bfseries}{\thesubsubsubsection}{1em}{}
\titlespacing*{\subsubsubsection}
{0pt}{3.25ex plus 1ex minus .2ex}{1.5ex plus .2ex}

\begin{document}

\title[Cosmological reconstructed solutions...]{
Cosmological reconstructed solutions in extended teleparallel gravity theories with a teleparallel Gauss-Bonnet term}
\author{\'Alvaro de la Cruz-Dombriz$^{1,2}$, Gabriel Farrugia$^{3,4}$, Jackson Levi Said$^{3,4}$ and Diego S\'aez-Chill\'on G\'omez$^{5}$}
\address{$^{1}$ Cosmology and Gravity Group, University of Cape Town, 7701 Rondebosch, Cape Town, South Africa}
\address{$^{2}$ Department of Mathematics and Applied Mathematics, University of Cape Town, 7701 Rondebosch, Cape Town, South Africa}
\address{$^{3}$ Department of Physics, University of Malta, Msida, MSD 2080, Malta}
\address{$^{4}$ Institute of Space Sciences and Astronomy, University of Malta, Msida, MSD 2080, Malta}
\address{$^{5}$ Institut de Ci\`{e}ncies de l'Espai, ICE/CSIC-IEEC, Campus UAB, Carrer de Can Magrans s/n, 08193 Bellaterra (Barcelona), Spain}


\begin{abstract}
In the context of extended Teleparallel gravity theories with a 3+1 dimensions Gauss-Bonnet analog term, we address the possibility of these theories reproducing several well-known cosmological solutions. In particular when applied to a Friedmann-Lema\^itre-Robertson-Walker geometry in four-dimensional spacetime with standard fluids exclusively. 
We study different types of gravitational Lagrangians and reconstruct solutions provided by analytical expressions for either the cosmological scale factor or the Hubble parameter. We also show that it is possible to find Lagrangians of this type without a cosmological constant such that the behaviour of the $\Lambda$CDM model is precisely mimicked. The new Lagrangians may also lead to other phenomenological consequences opening up the possibility for  new theories to compete directly with other extensions of General Relativity.
\end{abstract}

\maketitle

\section{Introduction}
\label{S1}
Teleparallel gravity is a gauge theory of the translation group, constructed by associating a Minkowskian tangent space to every point of the spacetime through the mathematical objects called vierbeins (also dubbed tetrads fields). The connection  
is assumed to be the so-called Weitzenb\"ock connection instead of the usual Levi-Civita connection. This choice leads to the absolute parallelism condition, meaning that the vierbeins 
are parallelly transported, a fact that 
explains the name of the theory ({\it c.f.} \cite{Teleparallelism} for a comprehensive review as well as \cite{Hayashi:1979qx} for further insight). Unlike the Levi-Civita connection, the Weitzenb\"ock connection is not commutative under the exchange of the lower indices, which not only induces a non-zero torsion, but also a zero Ricci scalar.  Thus a non-vanishing torsion $T$ emerges in a natural way at the level of the action. Originally proposed as an attempt at unifying gravitation and electromagnetism by Einstein, the Teleparallel Lagrangian is linear in $T$ and turns out to be an alternative description of  General Relativity (GR), such that every solution of GR is also a solution for the so-called Teleparallel Equivalent of General Relativity (TEGR). 

In analogy of higher-order theories of gravity, such as fourth-order $f(R)$ theories~\cite{fR_varia}--\cite{Cognola:2007zu},
 TEGR has been extended by constructing gravitational Lagrangians in terms of more complex functions of the torsion scalar in order to determine its relevance in cosmological contexts. Such extensions have been referred to as $f(T)$ theories (for a review see \cite{Cai:2015emx}) and were introduced for the first time in the literature as an alternative to the inflaton \cite{Ferraro:2006jd}, but were rapidly proposed also as a dark energy candidate \cite{Bengochea:2008gz}, and its cosmological properties extensively studied \cite{Cai:2011tc}--\cite{Odintsov:2015uca}. 
However, when introducing extra terms in the action, the aforementioned equivalence between GR and TEGR now does not remain between $f(T)$ and $f(R)$ theories. For instance,  the $f(T)$ gravitational field equations are second order whereas extensions of GR, such as $f(R)$ theories, usually host higher-order field equations. The equivalence between GR and TEGR at the level of field equations is retained due to boundary term difference between $R$ and $T$ not contributing to the latter. However, extending the theories to arbitrary functions of curvature or torsion generates differences in the field equations dependent on this boundary term difference, deviating from the equivalence between the corresponding theories \cite{Bahamonde:2015zma}. This results in interesting differences, such as the absence of extra gravitational-wave modes when compared to GR  \cite{Bamba:2013ooa}. \\


%
%

As a main drawback, it is widely known that extensions of TEGR, such as $f(T)$ gravity are not invariant under local Lorentz transformations within the context of absolute parallelism~\cite{Li:2010cg}. Thus, the field equations will be sensitive to the choice of tetrads \cite{good-and-bad}. The key-point lies in performing a correct parallelisation of spacetime, in order to obtain solutions of the field equations capable of, for instance, recovering the desired metric, such as Friedmann-Lema\^itre-Robertson-Walker (FLRW) geometries \cite{ref27}, the vacuum static and spherically symmetric (Schwarzschild) solution 
~\cite{Meng:2011ne}--\cite{Ferraro:2011ks}, or testing the validity of Birkhoff's theorem~\cite{Meng:2011ne, Tamanini:2012hg, Boehmer:2011gw}. However, this parallelisation procedure is not usually sufficient to keep the same predictions as GR, for instance junction conditions become stricter and depend upon both the $f(T)$ model and the choice of the tetrads, but the theory can contain the usual results of TEGR, as the Oppenheimer-Snyder collapse scenario by the appropriate choice of tetrads~\cite{delaCruz-Dombriz:2014zaa}. More recently,  a general set of acceptable junction conditions for $f(T)$ gravity via the variational principle was obtained in \cite{Velay-Vitow:2017odc}. Moreover, in analogy to metric theories, conformal invariants Lagrangians can be also constructed in the framework of Teleparallel gravities \cite{Bamba:2013jqa}. Also the very existence of black-hole configurations in other theories with torsion has been subject of study \cite{Cembranos_varia}.  
\newline

By relaxing this condition of absolute parallelism, the covariant form of $f(T)$ gravity can be obtained \cite{Krssak:2015oua,Golovnev:2017dox}. Taking this approach, tetrads will have an associated quantity (known as the spin-connection) which accounts for any deficit in producing the field equations. In this work, we do observe the absolute parallelism condition.
\newline

Cosmological solutions in $f(T)$ Teleparallel theories have also been given extensive coverage in the literature  \cite{Bengochea:2008gz, Ferraro:2011us, Wu:2010mn}.  Analogously 
to the subtleties about the correct choice of tetrads in static scenarios,  the use of diagonal tetrads in spherical coordinates in $f(T)$ theories, constrains the gravitational Lagrangian, since $f_{TT}=0$ is a necessary condition to fulfill the field equations. In other words, the $f(T)$ Teleparallel gravity would reduce to GR. 
Nonetheless, rotation of these tetrads shows that the gravitational Lagrangian appears to be unconstrained and consequently, general classes of these models are a priori permitted \cite{good-and-bad}. In addition, observational tests on $f(T)$ gravities states its suitability for being considered a serious competitor to other more standard cosmological models. Other extensions of Teleparallel gravity as couplings between the energy-momentum tensor and the torsion at the level of the action show the same consistent cosmological behaviour when tested with observational data \cite{Saez-Gomez:2016wxb}.
%
\newline

In addition, model-independent techniques such as cosmography~\cite{cosmo1:2}, have become essential to guarantee that the statistical outcomes do not depend upon the choice of the model. In fact, few references have been addressing the competitiveness of cosmographic techniques as suitable for reconstructing theoretical models beyond GR \cite{cosmo9}.
For teleparallel gravity theories, authors in~\cite{Capozziello:2015rda} used cosmographic techniques to extract some constraints on the redshift transition determining the onset of cosmic acceleration on claimed viable $f(T)$ forms and performed a Monte Carlo fitting using supernovae data. More recently in \cite{JCAP_Cosmography_Reverberi} (see further details therein), a more general Monte Carlo analysis with a Metropolis-Hastings algorithm with a Gelman-Rubin convergence criterion
was performed by using Union 2.1 supernovae catalogue, baryonic acoustic oscillation data 
and $H(z)$ differential age compilations. In the aforementioned reference, the cosmographic study showed that all $f(T)$ model parameters were compatible with zero at about 1--$\sigma$ level in the all--$z$ analysis. Still, the 95\% confidence levels allow for quite a large parameter range, particularly for higher derivatives of the $f(T)$ Lagrangian. \\

With the same aim, another type of extension of Teleparallel gravity has been recently proposed, where the analog $T_G$ to the Gauss-Bonnet invariant $G$ is constructed in terms of the torsion tensor \cite{Kofinas:2014owa}. In what follows, we shall refer to this as the TEGB term. The resulted object just differs in a total derivative with respect to the Gauss-Bonnet invariant, similarly as the torsion tensor with respect to the Ricci scalar curvature, such that both objects are equivalent at the linear level in the action, where in 4-dimensions they become a total derivative. Notwithstanding, extensions beyond the linear order imply modifications in the field equations, which may provide new insights at the cosmological level, such that when considering a more general action $f(T,T_G)$ both theories become different as in the case of $f(R)$ and $f(T)$, and also different from other extensions of Teleparallel gravity as $f(T)$ gravity. The resulting action is a new theory with new properties that may also reproduce suitable cosmic expansion histories. In this sense, some Lagrangians of this type have been studied aiming to unify the inflationary epoch with the dark energy era by quintessence or phantom-like fluids \cite{Kofinas:2014daa,Bahamonde:2016kba}. Also the phase space of such Lagrangians have been analysed leading to a wide range of possibilities \cite{Kofinas:2014aka}. The Noether Symmetry Approach has been also applied to  $f(T,T_G)$ gravities which helps to fix the form of the action \cite{Capozziello:2016eaz}.\\

In this paper, we shall thoroughly analyse $f(T,T_G)$ gravities in the framework of spatially flat FLRW cosmologies. 
We shall consider well-known cosmological solutions whose form is given by either the cosmological scale factor or the Hubble parameter. Essentially, we shall study power-law solutions whose correspondence in $f(R)$ fourth-order gravity theories is well-known \cite{Nojiri:2003ft} as well as in $f(R,G)$ Gauss-Bonnet gravities \cite{Nojiri:2005vv,Calcagni:2006ye}. Moreover, a $\Lambda$CDM cosmological evolution is also considered with the aim of being reconstructed by this kind of $f(T, T_G)$ Lagrangians without a cosmological constant term. Analogous Lagrangians were obtained for $f(R)$ gravity \cite{delaCruzDombriz:2006fj} and  $f(R,G)$ gravity \cite{Elizalde:2010jx}. 
Thus, in the following sections we shall address the degeneracy problem of these theories when compared to the Einsteinian relativity in a cosmological homogeneous and isotropic spacetime. 
These mimicked features may be considered as either an advantage or as an inconvenience. Anyhow, cosmological models within a class of gravitational theories capable of recovering the $\Lambda$CDM Concordance Model expansion history in different cosmological eras 
would indeed deserve further scrutiny. \\
\newline

The paper is organised as follows: in Section \ref{S2} we shall present the general features of the $f(T,T_G)$ gravity theories 
 by writing the basic equations which will be thoroughly studied in the following sections and we shall briefly discuss the tetrad choice subtleties. 
 In Section \ref{S3} we shall discuss the reconstruction of well-known cosmological scale factor solution for several paradigmatic classes - both additive and multiplicative form - of these theories.  
 Then Section \ref{S4} addresses the same issue when the desired quantity to reconstruct is the Hubble parameter by including de Sitter 
 scenario.
 Finally, Section \ref{S5} is devoted to study the possibility of reconstruction for the $\Lambda$CDM (dust and cosmological constant) solution as provided in GR. We shall prove that suitable $f(T,T_G)$ Lagrangians with no cosmological constant can be found, being such Lagrangians capable of hosting both Minkowski and Schwarzschild as vacuum solutions.
We conclude the paper by giving our conclusions in Section \ref{S8}. Two appendices, \ref{AppendixA} and \ref{AppendixB}, are included at the end of the communication to provide
explicitly the limiting cases on the torsion scalar when the reconstructed Lagrangians involve transcendental hypergeometric functions.
\newline

Throughout the paper we shall follow the following conventions: the Weitzenb\"ock connection as defined in the following Section will be denoted by $\tilde{\Gamma}^{\alpha}_{\mu\nu}$. $D_{\mu}$ shall represent
 the covariant derivative with respect to the Levi-Civita connection $\Gamma^{\alpha}_{\mu\nu}$.
%
Greek indices such as $\mu, \nu...$ shall refer to spacetime indices whereas latin letters $a, b, c...$ refer to the tetrads indices associated to the tangent space. 

\section{$f(T,T_G)$ theories}
\label{S2}

TEGR theories ({\it c.f.} \cite{Teleparallelism} for the rudiments of theories with torsion) and its extensions are constructed starting by defining the objects known as vierbeins  $e_{a}(x^{\mu})$, such that
\be
{\rm d}x^{\mu}=e_{a}^{\;\;\mu}\omega^{a}\; , \quad \omega^{a}=e^{a}_{\;\;\mu}{\rm d}x^{\mu}\; ,
\label{1.1}
\ee
which relate the tangent space at every point $x^{\mu}$ to the spacetime of a manifold described by the metric:
\begin{eqnarray}
{\rm d}s^{2} &=&g_{\mu\nu}{\rm d}x^{\mu}{\rm d}x^{\nu}=\eta_{ab}\omega^{a}\omega^{b}\label{1}\; ,
\label{1.1a}
\end{eqnarray} 
where $\eta_{ab}={\rm diag}(-1,1,1,1)$ holds for the Minkowskian metric. In addition, the tetrads accomplish the following properties:
\be
 e_{a}^{\;\;\mu}e^{a}_{\;\;\nu}=\delta^{\mu}_{\nu}\ , \quad e_{a}^{\;\;\mu}e^{b}_{\;\;\mu}=\delta^{b}_{a}\ .
 \ee
As in any gravitational theory, a connection provides the way of defining the covariant derivatives. In the case of Teleparallel gravity, the connection is constructed based on the translation group, which parallelly transports the tetrads, the so-called absolute parallelism condition, and is given by the Weitzenb\"{o}ck connection, 
\begin{eqnarray}
\tilde{\Gamma}^{\alpha}_{\mu\nu}=e_{a}^{\;\;\alpha}\partial_{\nu}e^{a}_{\;\;\mu}=-e^{a}_{\;\;\mu}\partial_{\nu}e_{a}^{\;\;\alpha}\label{co}\; ,
\label{WC}
\end{eqnarray}
Unlike the Levi-Civita connection, the Weitzenb\"{o}ck connection is not symmetric leading to a non-vanishing torsion tensor,
\begin{eqnarray}
T^{\alpha}_{\;\;\mu\nu}&=&\tilde{\Gamma}^{\alpha}_{\mu\nu}-\tilde{\Gamma}^{\alpha}_{\nu\mu}=e_{a}^{\;\;\alpha}\left(\partial_{\nu} e^{a}_{\;\;\mu}-\partial_{\mu} e^{a}_{\;\;\nu}\right)\ . 
\label{tor}
\end{eqnarray}
Whereas, the Riemann tensor for such a connection becomes null
$R^{\mu}_{\;\;\lambda\nu\rho}\left(\tilde{\Gamma}\right)=0\ .$
As well known, every connection can be expressed in terms of the Levi-Civita connection and its antisymmetric part, so the difference between the Weitzenb\"ock  and the Levi-Civita connection gives rise to the so-called contorsion tensor,
\be
K^{\alpha}_{\;\; \mu\nu}= \tilde{\Gamma}^{\alpha}_{\mu\nu}-\Gamma^{\alpha}_{\mu\nu}=\frac{1}{2}\left(T_{\mu\;\;\ \nu}^{\;\; \alpha}+T_{\nu\;\;\ \mu}^{\;\; \alpha}-T_{\;\; \mu \nu}^{\alpha}\right)\;,
\label{contor2}
\ee
whereas we can also define the superpotential:
\begin{eqnarray}
S_{\alpha}^{\;\;\mu\nu}&=&\frac{1}{2}\left( K_{\;\;\;\;\alpha}^{\mu\nu}+\delta^{\mu}_{\alpha}T^{\beta\nu}_{\;\;\;\;\beta}-\delta^{\nu}_{\alpha}T^{\beta\mu}_{\;\;\;\;\beta}\right)\label{s}\;,
\end{eqnarray}
whose contraction with the torsion tensor \eqref{tor} leads to the torsion scalar,
\begin{eqnarray}
T=T^{\alpha}_{\;\;\mu\nu}S^{\;\;\mu\nu}_{\alpha}=\frac{1}{4}T^{\lambda}_{\;\;\;\mu\nu}T_{\lambda}^{\;\;\;\mu\nu}+\frac{1}{2}T^{\lambda}_{\;\;\;\mu\nu}T_{\;\;\;\;\;\lambda}^{\nu\mu}-T^{\rho}_{\;\;\;\mu\rho}T_{\;\;\;\;\;\nu}^{\nu\mu}\, .
\label{scalar-torsion}
\end{eqnarray}
By the use of the contorsion tensor, one can easily achieve the relation among the Ricci scalar and the torsion scalar:
\be
R\,=\,-T-2D_{\mu}T^{\nu\mu}_{\;\;\;\;\;\nu}\ .
\label{RS}
\ee
Thus the gravitational action for TEGR is solely given by the torsion scalar (\ref{scalar-torsion}), 
\begin{eqnarray}
\label{action}
S_G=-\frac{1}{2\kappa^2}\int e\ T\ {\rm d}^4x\; ,
\label{TeleAction}
\end{eqnarray}
where $\kappa^2=8\pi G$ and $e=$ det $\left(e^{a}_{\;\;\mu}\right)$. It is straightforward to check that the above action is completely equivalent to the standard Einstein-Hilbert GR one at the level of field equations and up to a boundary term difference in the action. Nevertheless, any extension that includes a non-linear function of the torsion scalar will not be equivalent to its analog in curvature gravity, as the total derivative in Eq. (\ref{RS}) can not be dropped out. \\

Similarly to the expression (\ref{RS}), one can find the equivalent to the Gauss-Bonnet term in Weitzenb\"{o}ck geometry by using the above expressions. Indeed, the Gauss-Bonnet invariant is given by:
\be
G=R_{\mu\nu\lambda\sigma}R^{\mu\nu\lambda\sigma}-4R_{\mu\nu}R^{\mu\nu}+R^2\ .
\label{GBterm}
\ee
Hence, the torsion analog is related to the Gauss-Bonnet by \cite{Kofinas:2014owa}:
\be
G=T_G+B_G,
\label{TGGB}
\ee
where the second term is a total derivative\footnote{The explicit form of this boundary term is given in \cite{Bahamonde:2016kba} in Eq. (24).} while $T_G$ is defined in terms of the contorsion tensor as follows:
\begin{eqnarray}
T_G&=&\left(K^{\alpha}_{\;\;\gamma\beta}K^{\gamma\lambda}_{\;\;\;\;\rho}K^{\mu}_{\;\;\epsilon\sigma}K^{\epsilon\nu}_{\;\;\;\;\varphi} -2K^{\alpha\lambda}_{\;\;\;\;\beta}K^{\mu}_{\;\;\gamma\rho} K^{\gamma}_{\;\;\epsilon\sigma}K^{\epsilon\nu}_{\;\;\;\;\varphi}\right. \nn
&&+\left. 2K^{\alpha\lambda}_{\;\;\;\;\beta}K^{\mu}_{\;\;\gamma\rho}K^{\gamma\nu}_{\;\;\;\;\epsilon}K^{\epsilon}_{\;\;\sigma\varphi}+2K^{\alpha\lambda}_{\;\;\;\;\beta}K^{\mu}_{\;\;\gamma\rho}K^{\gamma\nu}_{\;\;\;\;\sigma,\varphi}\right)\delta^{\beta\rho\sigma\varphi}_{\alpha\lambda\mu\nu}.
\label{TG}
\end{eqnarray}
As in the case of the TEGR Lagrangian (\ref{TeleAction}), any action linear in $T_G$, by grace of \eqref{TGGB}, would be equivalent to a linear action containing the Gauss-Bonnet term and the total derivative $B_G$. However, any function with a non-linear dependence in either $G$ or $T_G$, would break the equivalence. This is precisely the class of theories to be studied in this paper, those whose total action (including the matter sector $S_m$) can be expressed as follows:
\begin{eqnarray}
S=S_G+S_m=\int e\left(\ f(T,T_G)\ +2\kappa^2\mathcal{L}_m\right){\rm d}^4x\; .
\label{fttGaction}
\end{eqnarray}
As can be found for instance in \cite{Kofinas:2014owa} the variation of the above action with respect to the metric tensor in the case of a spatially flat FLRW background renders the following two independent equations 
\begin{eqnarray}
& &f - 12H^2 f_T - T_G f_{T_G} +24H^3\dot{f}_{T_{G}}\,=\,2\kappa^2\rho_m\,, \label{Friedmann1} \\ 
& &f-4\left(3H^2+\dot{H}\right)f_T-4H\dot{f}_{T}-T_{G} f_{T_{G}}+\frac{2}{3H}T_{G} \dot{f}_{T_{G}}+8H^2\ddot{f}_{T_{G}}\,=\,-2\kappa^2p_m\,, \label{Friedmann2} \nonumber\\
&&
\end{eqnarray}
in the presence of a perfect fluid\footnote{The standard energy-momentum tensor definition $\mathcal{T}^{\;\nu}_{\mu}=\frac{e_{a}^{\;\;\nu}}{e}\frac{\delta\mathcal{L}_\mathrm{m}}{\delta e_{a}^{\;\;\mu}}$  has been used with $\mathcal{T}^{\;0}_{0}=\rho_m$ and $\mathcal{T}^{\;i}_{i}=-p_m$ ($i=1,2,3$). Note that this fluid is covariantly conserved as the left-hand side (l.h.s.) of the $f(T,T_G)$ field equations can be shown to be indeed covariantly conserved \cite{Kofinas:2014owa}.} of density $\rho_m$ and pressure $p_m$, with $H\equiv \dot{a}/a$ being $a$ the cosmological scale factor, dot denotes derivative with respect to the cosmic time $t$ and subindices $T$ and $T_G$ stand for derivatives of $f$ with respect to the corresponding argument. For FLRW geometries, one can show that $T$ and $T_G$ can be expressed in terms of the Hubble parameter and its first time derivative as follows
\begin{eqnarray}
T=6H^2\;\;\;;\;\;\; T_G=24H^2(\dot{H}+H^2)\,.
\label{T_TG_expressions}
\end{eqnarray}
In the case of a flat FLRW metric, $T_G$ coincides with its GR counterpart, $G$. The Lagrangian $f(T,T_G)$ for this spacetime is equivalent at the level of equations to the Lagrangian of the form $f(T,G)$. Thus solving the above field equation\footnote{ It is precisely the fact that the field equations are covariantly conserved which allows us to solely consider Eq. \eqref{Friedmann1} together with the energy-momentum conservation in order to perform our reconstruction techniques.} \eqref{Friedmann1} for general functions $f$ proves to be sufficiently cumbersome; however, solutions can be extracted when classes of models for $f$ with a given dependence in both $T$ and $T_G$ are considered.  Thus, in the next sections such classes of gravitational Lagrangians $f(T,T_G)$ would be reconstructed by assuming some paradigmatic cosmological solutions for either the cosmological scale factor or the Hubble parameter.  In addition, we require our Lagrangian to be able to recover cosmological vacuum solutions, i.e., we require that under the non existence of neither matter nor pressure, both $T$ and $T_G$ would be null. By taking a look to the trace equation as derived from the usual combination of Eqs. \eqref{Friedmann1} and \eqref{Friedmann2}, one obtains that the sufficient condition turns out to be $f(0,0) = 0$.

\section{Reconstruction through the scale factor}
\label{S3}

In this section, we investigate the possibility of obtaining suitable gravitational Lagrangians $f(T,T_G)$ capable of reproducing the cosmological evolution as provided by the cosmological scale factor. 
%
Due to its interest, we shall consider a power-law scale factor expressed in terms of the cosmic time $t$. In other words, we assume that $a(t) \propto t^{\alpha}$ where $\alpha \neq 0$ is some constant ($\alpha = 0$ corresponds to a static universe which is not of interest here). This type of reconstruction has been studied in a number of other extensions of GR such as Refs.\cite{JCAP_Cosmography_Reverberi,delaCruzDombriz:2011wn,Makarenko:2012gm,Bamba:2014mya}, among others. Moreover, the power law scale factor has shown promise for explaining the late time evolution of the universe \cite{Sethi:2005au,Kaeonikhom:2010vq,Ahmad:2013ygu}. \newline

For this kind of cosmic evolution, $H$, $T$ and $T_G$ take the following forms,
\begin{align}\label{eq:parameters-powerlawscale}
H &= \dfrac{\alpha}{t}, & T &= 6\dfrac{\alpha^2}{t^2}, & T_G &= 24\dfrac{\alpha^3}{t^4}\left(\alpha-1\right) = \dfrac{2}{3}\left(1-\dfrac{1}{\alpha}\right) T^2. 
\end{align}
Here, $T_G$ can be either positive, zero or negative depending on the value of $\alpha$, which can be characterised as follows:
\begin{itemize}
\item $T_G = 0$ when $\alpha \in\{0,\, 1\}$. However, the $\alpha = 0$ case is neglected. The case $\alpha=1$
is important since it enables us to find the TEGB terms able to explain the transitioning point towards an accelerated expansion. Indeed, the deceleration parameter $q \equiv - a\ddot{a}/\dot{a}^2$ is given by
\begin{equation}\label{eq:dec-powerscale}
q = -1+\dfrac{1}{\alpha}\,,
\end{equation}
for power-law scale factors. For $\alpha = 1$, $q = 0$, thus representing this transitioning point. 

\item $T_G > 0$ when $\alpha < 0$ or $\alpha > 1$.

\item $T_G < 0$ when $0 < \alpha < 1$.
\end{itemize}
These relations above will prove to be useful when examining the some cosmological solutions below.
Throughout this section, perfect fluids with a constant equation of state $w\equiv p_m/\rho_m$ are considered. Then, the Friedmann equation \eqref{Friedmann1} becomes
\begin{equation}
f - 2T f_T - T_G f_{T_G} - \dfrac{8\alpha T^2}{t^2} f_{TT_G} - \left(\dfrac{48\alpha^3}{t^4}\right)^2 (\alpha-1) f_{T_{G}T_G} = T_0 \Omega_{w,0} a^{-3(1+w)}\,.
\end{equation} 
Here  $\Omega_{w,0}\equiv 2\kappa^2\rho_{m,0}/3H_0^2$, $T_0 \equiv 6H_0^2$ and $H_0=H(t_0)$, where $t_0$ represents the cosmic time today. We will follow this notation along the paper. Let us now consider several paradigmatic functional forms of $f(T, T_G)$ in order to determine which gravitational Lagrangians are able to reproduce power-law scale factors. 

\subsection{$f(T,T_G) = g(T) + h(T_G)$}
\label{3.1}
Whenever the Lagrangian can be expressed as a sum of two functions $g$ and $h$, each being dependent on $T$ and $T_G$ respectively, the Friedmann equation \eqref{Friedmann1} reduces to
\begin{equation}
g + h - 2T g_T - T_G h_{T_G} + 24 H^3 h_{T_{G}T_G} \dot{T}_G = T_0 \Omega_{w,0} a^{-3(1+w)}.
\label{Eq_scalefactor_additive}
\end{equation} 
The scale factor on the right-hand side (r.h.s.) can be represented either in terms of $T$ or $T_G$ only\footnote{The cosmic time is expressed in terms of $T$ or $T_G$ since the relation \eqref{eq:parameters-powerlawscale}  is invertible.}. 
Although the choice of representation is arbitrary, in TEGR the fluid is usually described by the torsion scalar $T$ so we have decided, without any loss of generality, that the scale factor be represented by $T$. Equivalently, the same particular solution can be obtained if the dynamical variable of the system is taken as, $T_G$, instead of, $T$. This can be achieved by the relationship shown in Eq.(\ref{eq:parameters-powerlawscale}) where the direct relation between these two quantities is explicitly presented. The difference between these two approaches takes hold when other epochs, or other evolution ansatz, are assumed.

In this way, the Eq. \eqref{Eq_scalefactor_additive} can be expressed in terms of a single variable on each side of the equation, i.e., $X(T) = Y(T_G)$ for some functions $X$ and $Y$. Since herein both $T$ and $T_G$ are treated as independent variables, the equality would only be valid provided that $X(T) = K = Y(T_G)$, where K is a constant. However, the particular solutions for $g$ and $h$ would lead just to a constant contribution $g_\text{part.}(T) = -h_\text{part.}(T_G) = K$, which cancel each other in the action. Hence, we explore the solutions of the homogeneous equations. In other words, from Eq. \eqref{Eq_scalefactor_additive}, one obtains two independent differential equations\footnote{This approach has been used in $f(R,G)$ gravity in Refs. \cite{Elizalde:2010jx, delaCruzDombriz:2011wn}.}, namely 
\begin{align}
& g - 2T g_T = T_0 \Omega_{w,0} \left(\dfrac{T_0}{T}\right)^{-\frac{3(1+w)\alpha}{2}}, \label{eq:power-add-gODE}\\
& h - T_G h_{T_G} - \dfrac{4{T_G}^2}{\alpha-1} h_{T_{G}T_G} = 0. \label{eq:power-add-hODE1}
\end{align}
We remark that when $\alpha = 1$, the ODE for $h(T_G)$ does not diverge since the coefficient of $h_{T_G T_G}$ is well-defined for all values of $\alpha$. In fact, the resulting limit is
\begin{equation}
\dfrac{4{T_G}^2}{\alpha-1}\bigg|_{\alpha = 1} = 2304\dfrac{\alpha^6}{t^8}\left(\alpha-1\right)\bigg|_{\alpha = 1} = 0.
\end{equation}

{\bf Resolution for $h(T_G)$}: In the event of $\alpha=1$, since $T_G = 0$ according to Eq. \eqref{eq:parameters-powerlawscale}, the $h$ function does not physically contribute to the evolution since $T_G$ is identically zero and any evolution will be solely generated by $g$ (since $T \neq 0$). For this case, the condition to fulfill would be that $h(0) = 0$ provided that both $h_{T_G}(0)$ and $h_{T_G T_G}(0)$ are finite. Under these conditions, a class of functions can be allowed to satisfy the condition $h(0) = 0$, for example $h \propto {T_G}^n$ for $n > 2$.

%

On the other hand, for values $\alpha \neq 1$, and making use of Eq. \eqref{eq:parameters-powerlawscale},
Eq. \eqref{eq:power-add-hODE1} can be fully expressed in terms of $T_G$ 
whose general solution is
\begin{equation}
h(T_G) = c_1\,{T_G}^{\frac{1-\alpha}{4}}  + c_2\,T_G \,,
\label{solution_h_alphaneq1}
\end{equation}
where $c_{1,2}$ are integration constants. In the last expression, the second term represents the Gauss-Bonnet term, which is a total derivative and can be removed from the Lagrangian.
\newline
 
In order to keep vacuum solutions, where $T=0$ and $T_G=0$, and keeping in mind the form of $h(T_G)$ given in (\ref{solution_h_alphaneq1}), the power of $T_G$ has to be non-negative i.e., $\alpha < 1$ unless $c_1 = 0$. The latter leading to a trivial solution of the equation (\ref{eq:power-add-hODE1}).
%
\\
\newline
{\bf Resolution for {\bf $g(T)$}}: in order to solve Eq. \eqref{eq:power-add-gODE} for the function $g(T)$, two cases are considered, namely 
\begin{itemize}
\item $1-3 \alpha (1+w) \neq 0$:
for such models, the function of $g(T)$ is given by
\begin{equation}
g(T) = c_3 \sqrt{T}+\frac{\Omega_{w,0} T_0}{1-3 \alpha (1+w)} \left(\frac{T_0}{T}\right)^{-\frac{3 \alpha (1+w)}{2}},
\label{scalefactor_solution_g_additive}
\end{equation}
where $c_3$ is an integration constant. The first term provides the same cosmological evolution that arises in Dvali, Gabadadze and Porrati (DGP) gravity ({\it c.f.} \cite{Linder:2010py, Dvali:2000hr} for further details). In order to recover vacuum solutions, we shall require $g(0) = 0$. This imposes the following condition, 
\begin{equation}
\alpha (1+w) > 0,
\label{condition_g_additive}
\end{equation}
where by hypothesis the case $\alpha(1+w) = 1/3$ (i.e., $w \neq -1 + 1/3\alpha$) is excluded. If the function $h$ is non-trivially null, condition \eqref{condition_g_additive} can be combined with the previously found condition $\alpha \leq 1$, which was shown to be necessary for non-trivial $h$ functions. Thus, by combining such two conditions the values of $\alpha$ and $w$ are constrained as follows
\begin{itemize}
\item[a)] for $\alpha < 0$, $w < -1$;
\item[b)] for $0 < \alpha \leq 1$, $w > -1$. 
\end{itemize} 
Recalling the deceleration parameter as given by Eq. \eqref{eq:dec-powerscale}, 
cosmological acceleration shows up provided $\alpha < 0$ and deceleration provided $0 < \alpha < 1$. \\ 

In GR (or equivalently TEGR) deceleration occurs with perfect fluids whose EoS parameter is given by $w>-1/3$. Nonetheless, our analysis in this section shows that the deceleration condition includes values of $w \leq -1/3$, such that the extra gravitational terms in the action are the responsible for that. 
Indeed, in the event that TEGR is to be recovered, then the exponent in the second term of Eq. \eqref{scalefactor_solution_g_additive} must be equal to $-1$, resulting into the condition $\alpha(1+w) = 2/3$, leading to 
\begin{equation}
w = -1+\dfrac{2}{3\alpha}.
\end{equation}

\item Finally, for the particular case $1-3 \alpha (1+w) = 0$, i.e.,
\begin{equation}
\alpha (1+w) = \dfrac{1}{3}\,,
\label{Relation_w_alpha_2}
\end{equation}
the function $g(T)$ which is the solution of Eq. \eqref{scalefactor_solution_g_additive} becomes
\begin{equation}\label{scalefactor_solution_g_additive_equal}
g(T) = c_4 \sqrt{T}-\frac{1}{2} \Omega_{w,0} T \sqrt{\frac{T_0}{T}} \ln \left(\dfrac{T_0}{T}\right),
\end{equation}
where $c_4$ is an integration constant. In this case, the vacuum solution $g(0)=0$ is recovered in the limit $T \rightarrow 0$ according to 
Eq. \eqref{Relation_w_alpha_2} and provided non-trivial solutions of $h$, i.e., the condition $\alpha \leq 1$ must be accomplished. Thus, in this scenario the possible values of $w$ would range from $w > -2/3$ and $w < -1$.
In terms of the deceleration parameter, during an acceleration phase $(\alpha < 0)$, we find $w < -1$ whilst during a decelerating phase $(0 < \alpha < 1)$, we find $w > -2/3$, such that the transition point occurs at $w = -2/3$. \newline
Here and in several of the models that follow, TEGR cannot be completely recovered. This is not uncommon in the more exotic theories of gravity. While the expansion history is constrained by the Hubble parameter in Eq.(\ref{eq:parameters-powerlawscale}) it would be interesting to investigate whether or how other cosmological properties may be physically different from the TEGR case but this is beyond the scope of this study.
\end{itemize}

The results of Section \ref{3.1} are summarised on Table \ref{Table1}.


\subsection{$f(T,T_G) = T g(T_G)$}
\label{3.2}

The next type of $f(T,T_G)$ models investigates the possibility of performing a rescaling on TEGR 
through a function $g(T_G)$ in terms of the Gauss-Bonnet term. For this model, the Friedmann equation reduces to
\begin{equation}
g + T_G g_{T_G} + \dfrac{48 \alpha^3}{t^4} \left(g_{T_G} + 2T_G g_{T_{G}T_G} \right) = - \Omega_{w,0} \left(\dfrac{t}{t_0}\right)^{-3(1+w)\alpha + 2}.
\label{rescaling_1_fTTG}
\end{equation} 

Analogously to the models studied in Section \ref{3.1} above, it is important to distinguish the $\alpha =1$ and $\alpha \neq 1$ cases.\\
\\
\textit{(i) \tab[0.5cm]} ${\alpha = 1}$:  Since $T_G = 0$ in such instances, this sets $g(T_G) = g(0)$, hence a constant. Therefore, the Lagrangian behaves as a rescaled TEGR by a suitable constant $g(0)$. Thus, the Friedmann equation \eqref{rescaling_1_fTTG} reduces to
\begin{equation}
g + T_G g_{T_G} + \dfrac{48}{t^4} \left(g_{T_G} + 2T_G g_{T_{G}T_G} \right)\bigg|_{T_G = 0} = - \Omega_{w,0} \left(\dfrac{t}{t_0}\right)^{-3(1+w)\alpha+ 2}.
\label{diego1}
\end{equation} 
By evaluating $T_G$, this equation is essentially an equation of the form
\begin{equation}\label{eq:power-Trscl-alpha1}
\mu + \nu t^{-4} = \beta t^{-3(1+w)\alpha+2},
\end{equation}
where the symbols $\mu \equiv g + T_G g_{T_G}$, $\nu \equiv 48\left(g_{T_G} + 2T_G g_{T_{G}T_G} \right)$ and $\beta \equiv - \Omega_{w,0} (1/t_0)^{-3(1+w)+ 2}$  have been introduced. Note that both $\mu$ and $\nu$ are evaluated at $T_G = 0$, so are constants. 
Here we can distinguish two cases, depending on the value of $\alpha$ with respect to the EoS parameter $w$:
\begin{itemize}
\item $\alpha=\frac{2}{1+w}$. In this case, $\mu=0$ in (\ref{eq:power-Trscl-alpha1}), while $\nu=\beta$. Then, the most general solution becomes:
\be
g(T_G)=\sum_{n=1}^{i}\alpha_nT_G^n\ ,
\label{diego2}
\ee
where $\alpha_n$ are arbitrary constants except $\alpha_2=\beta/48$.
\item $\alpha=\frac{2}{3(1+w)}$. In this case, the time variable in the equation (\ref{eq:power-Trscl-alpha1}) is removed in the rhs, such that $\nu=0$ and $\mu=\beta$. Then, the following solution holds:
\be
g(T_G)=\beta+\sum_{n=2}^{i}\alpha_nT_G^n\ .
\label{diego3}
\ee
We remark that the TEGR contribution in the Lagrangian is obtained by setting $\beta = -1$, and requiring the parameters $\alpha_n$ to all vanish.
\end{itemize}
Note that both Lagrangians contain vacuum solution, since both are finite when evaluating at $T_G=0$.
\newline
\\
\textit{(ii) \tab[0.5cm]} ${\alpha \neq 1}$:  In this case, Eq. \eqref{rescaling_1_fTTG} can be expressed in terms of $T_G$ only
\begin{equation}
g + \dfrac{\alpha+1}{\alpha -1} T_G g_{T_G} + \dfrac{4 {T_G}^2}{\alpha -1} g_{T_{G}T_G} = - \Omega_{w,0} \left(\dfrac{T_{G,0}}{T_G}\right)^{\frac{-3(1+w)\alpha + 2}{4}}\,,
\label{rescaling_2_fTTG}
\end{equation}
whose solution becomes
\begin{align}
g(T_G) &= c_1 {T_G}^{m_{+}}+c_2 {T_G}^{m_{-}} +A\left(\frac{T_{G,0}}{T_G}\right)^{\frac{2-3\alpha (1+w)}{2}},
\label{sol_rescaling_2_fTTG}
\end{align}
where 
\bea
m_{\pm}=\frac{1}{8} \left(3-\alpha \pm\sqrt{\alpha^2-22\alpha+25}\right)\ , \nonumber \\
A=-\frac{2 (\alpha -1) \Omega_{w,0}}{3 [6 \alpha ^2 w^2+\left(13 \alpha ^2-11 \alpha \right) w+7 \alpha ^2-11 \alpha +4]}\ ,
\label{diego4}
\eea
and $c_{1,2}$ are integration constants provided that the denominator in (\ref{diego4}) is non-zero, i.e.,
\begin{equation}
w \neq \frac{11-13 \alpha \pm \sqrt{\alpha^2 -22 \alpha +25}}{12 \alpha }.
\end{equation}
It is also important to impose a real power in order to exclude complex Lagrangians (unless both $c_1$ and $c_2$ are zero). This is possible provided that the square-root of exponents in Eq.  \eqref{sol_rescaling_2_fTTG} first line are non-negative, i.e.,
\begin{equation}
\alpha^2-22\alpha+25 \geq 0 \implies \alpha \leq 11 - 4 \sqrt{6} \text{ or } \alpha\geq 11 + 4 \sqrt{6},
\label{alpha_values}
\end{equation}
or approximately $\alpha \lesssim 1.2$ or $\alpha \gtrsim 20.8$. \newline

The final issue to consider for solutions of the form \eqref{sol_rescaling_2_fTTG} is to recover vacuum solutions, i.e., $g(0)=0$. This is possible as far as the exponents in (\ref{sol_rescaling_2_fTTG}) lead to positive powers of $T_G$, which can be provided by setting $\alpha (1+w) > 0$ in the last term and fixing $c_{1,2}=0$ in case of $m_{\pm}<0$, which holds for the first term when $1 < \alpha \leq 11-4 \sqrt{6}$ $(1 < \alpha \lesssim 1.2)$ whilst for the second term, $\alpha \leq 11-4 \sqrt{6}$ $(\alpha \lesssim 1.2)$.\\



In terms of the expansion of the universe, we reach the following conclusions: if the $c_1$ term is retained, the values of $\alpha$ define an accelerating universe even with the presence of fluids with an EoS parameter $w \geq -1/3$. On the other hand, if the $c_2$ term is retained, $\alpha < 0$ is described by phantom fluids (which excludes the standard $-1 \leq w \leq -1/3$ domain) whilst for $0 < \alpha < 1$, which describes a decelerating universe, a non-phantom fluid is required (which also includes the standard $-1 < w \leq -1/3$ domain). \newline



\subsection{$f(T,T_G) = T_G g(T)$}
\label{3.3}

Let us now consider the class of gravitational Lagrangians rescaled by a function of the torsion scalar $T$. For this class of models, the Friedmann equation becomes
\begin{equation}
- \dfrac{4}{3}T^3 g_T = T_0 \Omega_{w,0} \left(\dfrac{T_0}{T}\right)^{-\frac{3(1+w)\alpha}{2}}\,,
\label{eq_scalepower-rescaling2}
\end{equation} 
whose solutions depend upon the relation between $\alpha$ and $w$, giving rise to the following cases:
%
%
\begin{enumerate}
\item {\bf $4-3\alpha(1+w) \neq 0$}: in this case, the solution of Eq. \eqref{eq_scalepower-rescaling2} is given by
\begin{equation}
g(T)=c_1+A\left(\frac{T}{T_0}\right)^{3\alpha(1+w)/2},
\label{g1_scalepower-rescaling2}
\end{equation}
where $A=\frac{3 \Omega_{w,0}}{2\left[4-3 \alpha (1+w)\right]} $ and $c_1$ is an integration constant. This constant corresponds to the Gauss-Bonnet term and hence can be removed. For vacuum solutions, by using the arguments in the previous section, the resulting power of $T$ in the last term of Eq. \eqref{g1_scalepower-rescaling2} has to be positive at all times. This leads to the condition
\begin{equation}
\alpha(1+w) > -\dfrac{2}{3}.
\end{equation}
excluding $\alpha(1+w) \neq 4/3$ since this leads to a singularity in the Lagrangian. No further constraints on $\alpha$ exist. \newline
\\
%
\item {\bf $4-3\alpha(1+w) = 0$}: in this case, the solution of Eq. \eqref{eq_scalepower-rescaling2} is given by
\begin{equation}\label{g1_scalepower-rescaling3}
g(T) = c_2+A \ln \left(\dfrac{T_0}{T}\right),
\end{equation}
where $A=\dfrac{3 \Omega_{w,0}}{4T_0}$  and $c_2$ is an integration constant. In this case, vacuum solutions exist and no further relation between $\alpha$ and $w$ is needed.

\end{enumerate}

\subsection{$f(T,T_G) = -T+T_G g(T)$}
\label{3.4}

The next form of $f$ considers again a rescaling of $T_G$ by a function in terms of the torsion scalar $T$, but this time including the effects of the TEGR term $-T$.  For this type of models, the Friedmann equation becomes
\begin{equation} \label{Eq_3.4}
T - \dfrac{4}{3}T^3 g_T = T_0 \Omega_{w,0} \left(\dfrac{T_0}{T}\right)^{-\frac{3(1+w)\alpha}{2}},
\end{equation} 
Analogously to the results in Sec. \ref{3.3}, the obtained solutions depend upon the relation between $\alpha$ and $w$. In fact solutions are identical to those obtained above in Sec. \ref{3.3}, except for the fact that solutions require an extra term which stems from the $T$ term on the l.h.s. of Eq. \eqref{Eq_3.4} being 
\begin{equation}\label{g1_scalepower-tegr-rescaling}
g_\text{part.}(T) = -\dfrac{3}{4 T}.
\end{equation}
Overall, according to the form of the Lagrangian this adds an extra contribution of $-T-\dfrac{3T_G}{4T}$ to the Lagrangian compared to solutions in Sec. \ref{3.3}. 


\subsection{$f(T,T_G) = -T+ \mu \left(\frac{T}{T_0}\right)^\beta \left(\frac{T_G}{T_{G,0}}\right)^\gamma $}
\label{sec_scalepower-productpowers}
Let us now assume a Lagrangian described by powers of the scalar functions together with the TEGR term to check whether such models are capable of reproducing power-law solutions. Here, $\mu$, $\beta$ and $\gamma$ are constants ($\mu$ with units of $T$ and $\beta$ and $\gamma$ dimensionless exponents). The Friedmann equation becomes 
\begin{equation}\label{Friedmann_powerlaw_productpowers}
T + \mu \left(\dfrac{T}{T_0}\right)^\beta \left(\dfrac{T_G}{T_{G,0}}\right)^\gamma \left[1- 2 \beta - \gamma - \dfrac{2\gamma}{\alpha -1} \left(\beta + 2 \gamma - 2\right) \right] = T_0 \Omega_{w,0} a^{-3(1+w)}\,,
\end{equation} 
where two cases deserve to be analised separately, namely
\begin{enumerate}
\item {\bf $\alpha = 1$}: In this case $T_G\equiv 0$ and the Lagrangian effectively reduces to TEGR. Thus the previous equation yields
\begin{equation}
T = T_0 \Omega_{w,0} a^{-3(1+w)}\,,
\end{equation} 
which, according to expression \eqref{eq:parameters-powerlawscale} fixes $w=-1/3$.
\item {\bf $\alpha \neq 1$}: In this case, the constant $\mu$ in Eq. \eqref{Friedmann_powerlaw_productpowers} can be found by evaluating the expression at current time
\begin{equation}
\mu = \dfrac{T_0 (\Omega_{w,0}-1)(\alpha-1)}{(\alpha-1)(1- 2 \beta - \gamma) - 2\gamma\left(\beta + 2 \gamma - 2\right)},
\label{mu_expression}
\end{equation}
provided that the denominator is non-zero (if this is the case, the Friedmann equation reduces to that of TEGR meaning that the power terms do not affect the power-law evolution). Thus, using 
expression \eqref{mu_expression}, the Friedmann equation can be expressed in a simpler form,
\begin{equation}
\dfrac{T}{T_0} + \left(\dfrac{T}{T_0}\right)^\beta \left(\dfrac{T_G}{T_{G,0}}\right)^\gamma  \left(\Omega_{w,0}-1\right) = \Omega_{w,0} a^{-3(1+w)}\,,
\end{equation}
which eventually can be rewritten in terms of cosmic time using \eqref{eq:parameters-powerlawscale} to become
\begin{equation}
\left(\dfrac{t}{t_0}\right)^{-2+3(1+w)\alpha} +  \left(\Omega_{w,0}-1\right)\left(\dfrac{t}{t_0}\right)^{-2\beta-4\gamma+3(1+w)\alpha} \,=\, \Omega_{w,0}.
\end{equation}

This leads to an important consequence. Since the RHS is constant in cosmic time, the l.h.s. must be time independent. This can only hold provided that the powers of $t$ are zero, which leads to the following constraints
\begin{align}
\beta+2\gamma &= 1, \label{eq_scalepower_powerlaw_cond1}\\
3\alpha(1+w) &= 2. \label{eq_scalepower_powerlaw_cond2}
\end{align}
The first condition constrains the relation between the exponents $\beta$ and $\gamma$ (this also ensures that vacuum solutions can be obtained) whilst the second one restricts the equation of state parameter of the fluid $w$ in relation to the power-law exponent $\alpha$. For matter and radiation dominated universes, the latter holds exactly as in GR ($w = 0$ and $\alpha = 2/3$ for matter whilst $w = 1/3$ and $\alpha = 1/2$ for radiation). Furthermore, coupled with the condition that the denominator of $\mu$ in Eq. \eqref{mu_expression} is non-zero, an extra condition is obtained
\begin{equation}\label{eq_scalepower_powerlaw_cond3}
\gamma \neq \dfrac{\alpha -1}{3\alpha -1}\,,
\end{equation}
which constrains the allowed values of $\gamma$. 

\end{enumerate}

The results of the whole Section \ref{S3} are summarised on Table \ref{Table1}. 

\begin{table}
\begin{tabular}{ | l | l | l |}
\hline
\hline
\multicolumn{3}{ |c| }{$f(T,T_G)=g(T)+h(T_G)$} \\
\hline
\hline
Parameters & \multicolumn{1}{ |c| }{$1-3\alpha(1+w)\neq 0$} & \multicolumn{1}{ |c| }{$1-3\alpha(1+w)=0$} \\ \hline
\multirow{2}{*}{$\alpha=1$} & \multicolumn{2}{ |c| }{$h(T_G)=0,\quad T_Gh'(T_G)=0, \quad T_Gh''(T_G)=0$ at $T_G=0$} \\ \cline{2-3}
& \multirow{2}{*}{$g(T)=c_1 \sqrt{T}+\frac{\Omega_{w,0} T_0}{1-3 \alpha (1+w)} \left(\frac{T_0}{T}\right)^{-\frac{3 \alpha (1+w)}{2}}$} & \multirow{2}{*}{$g(T)=c_1 \sqrt{T}-\frac{1}{2} \Omega_{w,0} T \sqrt{\frac{T_0}{T}} \ln \left(\dfrac{T_0}{T}\right)$}\\ \cline{1-1}
\multirow{3}{*}{$\alpha\neq1$} & & \\ \cline{2-3}
& \multicolumn{2}{ |c| }{\multirow{2}{*}{$h(T_G)=c_1{T_G}^{\frac{1-\alpha}{4}} +c_2 T_G$}} \\
& \multicolumn{2}{|c|}{} \\
\hline
\hline
\multicolumn{3}{c}{} \\
\hline
\hline
\multicolumn{3}{ |c| }{$f(T,T_G)=T g(T_G)$} \\
\hline
\hline
Parameters & \multicolumn{1}{ |c| }{$\alpha=-\frac{2}{1+w}$} & \multicolumn{1}{ |c| }{$\alpha=\frac{2}{3(1+w)}$} \\ \hline
\multirow{1}{*}{$\alpha=1$} & \multirow{1}{*}{$g(T_G)=\sum_{n=1}^{i}\alpha_nT_G^n$, $i \in \mathbb{Z}^+, i \geq 1$} & \multirow{1}{*}{$g(T_G)=\beta+\sum_{n=2}^{i}\alpha_nT_G^n$, $i \in \mathbb{Z}^+, i \geq 2$} \\ \cline{1-3}
\multirow{2}{*}{$\alpha\neq1$} & \multicolumn{2}{|c|}{\multirow{2}{*}{$g(T_G)=c_1T_G^{m_+}+c_2T_G^{m_-}+A(T_{G,0}/T_G)^{\frac{2-3\alpha(1+w)}{2}}$}} \\
& \multicolumn{2}{|c|}{}\\
\hline
\hline
\multicolumn{3}{c}{} \\
\hline
\hline
\multicolumn{3}{ |c| }{$f(T,T_G)=T_G g(T)$} \\
\hline
\hline
Parameters & \multicolumn{1}{ |c| }{$4-3\alpha(1+w)\neq0$} & \multicolumn{1}{ |c| }{$4-3\alpha(1+w)=0$} \\ \hline
\multirow{2}{*}{$\alpha\neq 1$} & \multirow{2}{*}{$g(T)=c_1+A(T/T_0)^{3\alpha(1+w)/2}$} & \multirow{2}{*}{$g(T)=c_1+A\ln\left(\frac{T_0}{T}\right)$} \\ 
& & \\
\hline
\hline
\multicolumn{3}{c}{} \\
\hline
\hline
\multicolumn{3}{ |c| }{$f(T,T_G)=-T+T_G g(T)$} \\
\hline
\hline
Parameters & \multicolumn{1}{ |c| }{$4-3\alpha(1+w)\neq0$} & \multicolumn{1}{ |c| }{$4-3\alpha(1+w)=0$} \\ \hline
\multirow{2}{*}{$\alpha\neq 1$} & \multirow{2}{*}{$g(T)=c_1+AT^{3\alpha(1+w)/2}-\frac{3}{4T}$} & \multirow{2}{*}{$g(T)=c_1+A\ln\left(\frac{T_0}{T}\right)-\frac{3}{4T}$} \\ 
& & \\
\hline
\hline
\multicolumn{3}{c}{} \\
\hline
\hline
\multicolumn{3}{ |c| }{$f(T,T_G) = -T+ \mu \left(\frac{T}{T_0}\right)^\beta \left(\frac{T_G}{T_{G,0}}\right)^\gamma $} \\
\hline
\hline
Parameters & \multicolumn{2}{ |c| }{} \\ \cline{1-1}
\multirow{2}{*}{$\alpha= 1$} & \multicolumn{2}{ |c| }{\multirow{2}{*}{Solution exists just for $w=-1/3$ with $\mu$, $\beta$ and $\gamma$ undetermined}} \\ 
& \multicolumn{2}{ |c| }{} \\
\hline
\multirow{3}{*}{$\alpha\neq 1$} & \multicolumn{2}{ |c| }{\multirow{3}{*}{$\mu = \dfrac{T_0 \left(\Omega_{w,0}-1\right)}{1- 2 \beta - \gamma - \dfrac{2\gamma}{\alpha -1} \left(\beta + 2 \gamma - 2\right)},
$ $\beta=1-\gamma$ and $\alpha=2/3(1+w)$}} \\ 
& \multicolumn{2}{ |c| }{} \\
& \multicolumn{2}{ |c| }{} \\
\hline
\hline
\end{tabular}
 \caption{Summary of the Lagrangians $f(T,T_G)$ that reproduce power law solutions of the type $a(t)\propto t^{\alpha}$. Note every gravitational action is subjected to the values of the free parameters, leading to analytical functions at every point (including vacuum) or to singular functions. Hence, depending on the free parameters, one action may be more convenient than others}
 \label{Table1}
\end{table}

\section{Reconstruction through the Hubble parameter}
\label{S4}
  
In this section, we consider reconstruction models for power-law solutions for the Hubble parameter, i.e., $H \propto t^\alpha$ for some real constant $\alpha$. Again, as in the previous case, there are numerous important samples of power-law ansatzs which can be taken for the Hubble parameter, some  of them widely studied in the literature, Refs.\cite{Barrow:1993zq,Jawad:2014kka,Pasqua:2015rxa,Khatua:2009zz}. 
Throughout this Section, $H$, $T$ and $T_G$ take the following forms,
\begin{align}
H &= H_0 \left(\dfrac{t}{t_0}\right)^\alpha, & T &= 6 H_0^2 \left(\dfrac{t}{t_0}\right)^{2\alpha}, & T_G &= 24H^3\left(\dfrac{\alpha}{t} + H\right).
\label{H_expressions}
\end{align}
Furthermore, the scale factor takes on two different forms depending on the value of $\alpha$. For $\alpha \neq -1$,
\begin{equation}\label{a_from_H}
a(t) = \exp \left\lbrace \dfrac{H_0 t_0}{1+\alpha}\left[\left(\frac{t}{t_0}\right)^{1+\alpha}-1\right]\right\rbrace,
\end{equation}
whilst for $\alpha = -1$,
\begin{equation}
a(t) = \left(\dfrac{t}{t_0}\right)^{H_0 t_0}.
\end{equation}
%
As in Sec. \ref{S3}, we now solve the Friedmann equation for the various models of $f(T,T_G)$ gravity. In the event that we impose $\alpha = -1$ then results presented in Sec. \ref{S3} are recovered. 
Thus, in what follows $\alpha \neq -1$ is assumed. In addition, the case of $\alpha \neq -1$ may correspond to the Little Rip scenario for $\alpha>0$, since the Hubble parameter turns out to be an increasing function of time tending to infinity at an infinite time. This scenario would increase the inertial force of the expansion, breaking down bound systems \cite{Frampton:2011sp}. This phenomenon has been previously considered in other modified gravity theories such as $f(R)$ gravity \cite{Nojiri:2011kd}.


\subsection{Reconstruction for de Sitter solutions}
\label{4.1} 
In the event of $\alpha=0$ in Eq. \eqref{H_expressions}, the Hubble parameter becomes constant, say $H = H_\text{dS}$. Therefore, $H$, $T$ and $T_G$ take the constant values,
\begin{align}
H &= H_\text{dS}, & T &= 6{H_\text{dS}}^2, & T_G &= 24{H_\text{dS}}^4. 
\label{deSitter_TTG}
\end{align}
%

In vacuum, both $T$ as $T_G$ become constant according to \eqref{deSitter_TTG}, so the Friedmann equation reduces to
\begin{eqnarray}
\label{Eq_vacuum_deSitter}
f - 2T f_T - T_G f_{T_G} = 0.
\end{eqnarray} 
Dealing this as a partial differential equation in $f$, this can be solved analytically to give 
\begin{eqnarray}
\label{Solution_deSitter}
f(T,T_G) = \sqrt{T}\, g\left(\dfrac{T_G}{\sqrt{T}}\right)\,,
\end{eqnarray} 
where $g$ is some arbitrary function.
Thus, an infinite class of solutions of the form \eqref{Solution_deSitter} exists depending upon the choice of the function $g$. As expected one finds that both the DGP term and Gauss-Bonnet terms are solutions for suitable functions $g$. 
Although an infinite class of solutions does indeed solve Eq. \eqref{Eq_vacuum_deSitter}, only a subset of these solutions provide vacuum solutions through the requirement $f(0,0) = 0$. For instance, the solution 
\begin{eqnarray}
f(T,T_G) = c_1 \sqrt{T} \left(\dfrac{T_G}{\sqrt{T}}\right)^{-1} = c_1 \dfrac{T}{T_G}\,,
\end{eqnarray}
where $c_1$ is an arbitrary constant, does not generate vacuum solutions. Since $T_G = 2T^2/3$ during de Sitter periods, $f \propto 1/T$ which does not give zero when $T \rightarrow 0$. Anyhow, suitable Lagrangians of the form \eqref{Solution_deSitter} can be found.


\subsection{$f(T,T_G) = g(T) + h(T_G)$}
\label{4.2}
For models which can be split as an addition of functions for $T$ and $T_G$, the Friedmann equation reduces to
\begin{equation}
g + h - 12H^2 g_T - T_G h_{T_G} + 24 H^3 h_{T_G T_G} \dot{T}_G = T_0 \Omega_{w,0} a^{-3(1+w)}.
\end{equation}
Since the scale factor can be expressed in terms of $T$ only, the equation can be separated into two decoupled differential equations for $g$ and $h$ (analogous to the case studied in 
Sec. \ref{3.1}). Thus the corresponding two differential equations are 
\begin{eqnarray}
&g - 2T g_T = T_0 \Omega_{w,0} \exp \left\lbrace\dfrac{3(1+w) H_0 t_0}{1+\alpha} \left[1-\left(\frac{T}{T_0}\right)^{\frac{1+\alpha}{2\alpha}}\right] \right\rbrace, \label{eq:powerH-add-gODE}\\
&h - T_G h_{T_G} + \alpha \dfrac{3\alpha -1 +4Ht}{(\alpha+Ht)^2} {T_G}^2 h_{T_G T_G} = 0. \label{eq:powerH-add-hODE}
\end{eqnarray}
\newline
\\
\textit{(i) \tab[0.5cm]} {\bf Finding $g(T)$}:  a general solution for $g$ for Eq. \eqref{eq:powerH-add-gODE} can not be found a priori. Nonetheless, an alternative consists of expressing the exponential function on the r.h.s in Eq.  \eqref{eq:powerH-add-gODE} as a power series. 
In other words, we solve the differential equation
\begin{equation}
g - 2T g_T = T_0 \Omega_{w,0} \sum\limits_{n=0}^\infty \dfrac{1}{n!}\left\{ \dfrac{3(1+w) H_0 t_0}{1+\alpha} \left[1-\left(\dfrac{T}{T_0}\right)^{\frac{1+\alpha}{2\alpha}}\right] \right\}^n
\end{equation}
Since the terms in the power-series are all continuous, the summation and integration can be exchanged, and hence a power-series analytical solution for $g$ can be generated, being
\begin{align}\label{solution_g_Add_Hubble}
g(T) &= c_1 \sqrt{T} \nonumber \\
&+ \Omega_{w,0} T_0 \sum\limits_{n=0}^\infty \frac{1}{n!} \left[\frac{3 H_0 t_0 (1+w)}{1+\alpha}\right]^n \, _2F_1\left[-\frac{\alpha}{1+\alpha},-n;\frac{1}{1+\alpha};\left(\frac{T}{T_0}\right)^{\frac{1+\alpha}{2 \alpha}}\right],
\end{align}
where $c_1$ is an integration constant and $_2F_1(a,b;c;z)$ is Gauss's hypergeometric function. Note that the first term in Eq. \eqref{solution_g_Add_Hubble} corresponds to the DGP term. A pertinent discussion now would be a careful study about the approximate form of the hypergeometric function appearing in solution \eqref{solution_g_Add_Hubble}. In order to do so, we have considered two limiting cases, $T/T_0 \ll 1$ and $T/T_0 \gg 1$, which are presented in \ref{AppendixA} and Table \ref{Table2}. These will be useful when studying the feasibility of vacuum solutions at the end of the Section.
\newline
\\
%
\textit{(ii) \tab[0.5cm]} {\bf Finding $h(T_G)$}: 
from \eqref{H_expressions}, one finds that 
\begin{equation}\label{TG_vs_t}
T_G = 24 {H_0}^3 \left(\dfrac{t}{t_0}\right)^{3\alpha} \dfrac{\alpha}{t}+ 24 {H_0}^4 \left(\dfrac{t}{t_0}\right)^{4\alpha} \equiv A t^{3\alpha -1} + B t^{4\alpha},
\end{equation}
where $A \equiv 24\alpha {H_0}^3 \left(\dfrac{1}{t_0}\right)^{3\alpha}$ and $B \equiv 24 {H_0}^4 \left(\dfrac{1}{t_0}\right)^{4\alpha}$. However, Eq. \eqref{TG_vs_t} is in general not invertible for $t$, i.e., there is no function $p$ such that $t = p(T_G)$, such that numerical resources become necessary. 
%
\\

{\bf About vacuum solutions}:  In order for the full gravitational Lagrangian $f(T,T_G) = g(T) + h(T_G)$ to 
recover vacuum solutions, i.e., $f(0,0) = 0$ there are two possible options: either $g(0) = -h(0)$ or $g(0) = h(0) = 0$. 
Since the argument of the hypergeometric function in Eq. \eqref{solution_g_Add_Hubble} is power dependent with exponent $\dfrac{1+\alpha}{2\alpha}$, we have two cases to consider for $g$ depending on the sign of this exponent.
Thus in order to deal with such a problem, let us study the solution \eqref{solution_g_Add_Hubble} and approximate solutions of Eq.  \eqref{eq:powerH-add-hODE} in the limit $T \rightarrow 0$.
 \newline

Let us first consider an order approximation. Since $Ht \sim t^{\alpha+1} \propto T^{\frac{1+\alpha}{2\alpha}}$, we have the following limits as $T \rightarrow 0$. For $\dfrac{1+\alpha}{2\alpha} > 0$ (i.e. $\alpha < -1$ or $\alpha > 0$), the $Ht$ terms go to zero whilst for $\dfrac{1+\alpha}{2\alpha} < 0$ (i.e. $-1 < \alpha < 0$), the $Ht$ terms go to infinity. Note that the case $\dfrac{1+\alpha}{2\alpha}  = 0$ is not considered here since this is true when $\alpha = -1$, which contradicts the assumption made in Eq. \eqref{a_from_H}. Thus in this limit, the coefficient of ${T_G}^2 h_{T_G T_G}$ in Eq. \eqref{eq:powerH-add-hODE} becomes
\begin{align}
\alpha \dfrac{3\alpha -1 +4Ht}{(\alpha+Ht)^2} &= \begin{cases}
\dfrac{3\alpha -1}{\alpha}, & \text{for } \alpha < -1 \text{ or } \alpha > 0,\\
        0, & \text{for } -1 < \alpha < 0.
\end{cases}
\end{align}
Thus, under this approximation $T \rightarrow 0$, Eq. \eqref{eq:powerH-add-hODE} can be indeed solved,
\begin{eqnarray}
h(T_G)\,&=&\,c_2\,T_G^{\frac{\alpha}{3\alpha-1}}+c_3\,T_G\,, \;\;\;\text{for} \;\;\; \alpha < -1 \text{ or } \alpha > 0\label{Sol_1_h_TG}\\
h(T_G)\,&=&\,c_4\,T_G\,, \;\;\;\;\;\;\;\;\;\;\;\;\;\;\;\;\;\;\;\;\text{for} \;\;\; -1<\alpha < 0 \text{ or } \alpha =1/3\label{Sol_2_h_TG}\,,
\end{eqnarray}
with $c_{2,3,4}$ arbitrary constants. The solution in Eq. \eqref{Sol_2_h_TG} corresponds to the standard boundary term, whereas for  solution \eqref{Sol_1_h_TG} the accomplishment of finite solutions at $T_G=0$ depend upon the values of $\alpha$ or alternatively the fixing of $c_1$ constant to zero.
%
%
Thus, since $h(0) = 0$ is assumed, we need that $g(0) = 0$ in Eq. \eqref{solution_g_Add_Hubble}.\\

Let us now study the effect of those two cases, namely $\dfrac{1+\alpha}{2\alpha}\gtrless0$ on the approximate solution for $g(T)$.
\begin{itemize}
\item For $\dfrac{1+\alpha}{2\alpha} > 0$ (i.e. $\alpha < -1$ or $\alpha > 0$), the hypergeometric function in the limit $T\rightarrow0$ becomes $_2F_1(a,b;c;0)$, which by definition is 1. Thus, the function $g(0)$ in Eq. \eqref{solution_g_Add_Hubble} becomes 
\begin{equation}
g(0) = \Omega_{w,0} T_0 \sum\limits_{n=0}^\infty \frac{1}{n!} \left[\frac{3 H_0 t_0 (1+w)}{1+\alpha}\right]^n = \Omega_{w,0} T_0 \exp \left[\frac{3 H_0 t_0 (1+w)}{1+\alpha}\right],
\end{equation}
which is non-zero. Thus, for this range of $\alpha$, the full gravitational Lagrangian $f(T,T_G)$ cannot host vacuum solutions. \newline

\item For $\dfrac{1+\alpha}{2\alpha} < 0$ (i.e. $-1 < \alpha < 0$), the argument in the hypergeometric function present in solution \eqref{solution_g_Add_Hubble} tends to infinity when $T\rightarrow0$. In this case, the hypergeometric function is zero. Thus, $g(0) = 0$ as needed and hence for this interval of values of parameter $\alpha$  vacuum solutions can be hosted. \newline

\end{itemize}
The results of Section \ref{4.2} as summarised on Table \ref{Table2}.

\subsection{$f(T,T_G) = T g(T_G)$}
\label{4.3}
For $T$ rescaling models, the Friedmann equation becomes
\begin{equation}\label{Eq_Sec_4.3}
g + T_G g_{T_G} - 4T\dot{T_G} H g_{T_G T_G} = -\dfrac{T_0}{T} \Omega_{w,0} \exp \left\lbrace\dfrac{3(1+w) H_0 t_0}{1+\alpha} \left[1-\left(\frac{T}{T_0}\right)^{\frac{1+\alpha}{2\alpha}}\right] \right\rbrace.
\end{equation} 
At this point, a number of problems arise. Firstly, the coefficient of $g_{T_G T_G}$ cannot be expressed in terms of $T_G$ only. This problem arises from the non-invertibility of the time parameter with $T_G$ similar to the scenario above in Sec. \ref{4.2}. This fact also implies that the r.h.s. of Eq. \eqref{Eq_Sec_4.3} cannot be expressed in terms of $T_G$. Thus, the resulting differential equation cannot be expressed in terms of the variable $T_G$ only and hence cannot be solved. Nonetheless, as described above the equation can be settled in terms of the time variable only. The resulting equation cannot be analytically solved thus one might consider solving these equations using numerical techniques. 
Furthermore, even if a solution is found, due to the problem of invertibility, the function $g$ would not be expressible in terms of $T_G$ analytically.

\subsection{$f(T,T_G) = T_G g(T)$}
\label{4.4}
For $T_G$ rescaling, the Friedmann equation becomes
\begin{equation}\label{Eq_Sec_4.4}
- \dfrac{4}{3} T^3 g_T = T_0 \Omega_{w,0}  \exp \left\lbrace\dfrac{3(1+w) H_0 t_0}{1+\alpha} \left[1-\left(\frac{T}{T_0}\right)^{\frac{1+\alpha}{2\alpha}}\right] \right\rbrace.
\end{equation} 
Analogously to the additive model in Sec. \ref{4.2}, finding the full analytical solution in Eq. \eqref{Eq_Sec_4.4} is not possible directly. However, the solution can be generated using power series. Expressing the exponential function as a power series, the Friedmann equation can be expressed as
\begin{equation}
- \dfrac{4T^3}{3} g_T = T_0 \Omega_{w,0} \sum\limits_{n=0}^\infty \dfrac{1}{n!}\left\{\dfrac{3(1+w) H_0 t_0}{1+\alpha}\left[1-\left(\dfrac{T}{T_0}\right)^{\frac{1+\alpha}{2\alpha}}\right]\right\}^n \,,
\end{equation} 
whose r.h.s., given the fact that $T>0$ according to \eqref{H_expressions},  is real at all times. Analogously to the additive model in Sec. \ref{4.1}, the terms in the power-series above are all continuous. Thus, the summation and integration can be exchanged leading to the following solution,
\begin{align}
&g(T) = c_1 \nonumber \\
&+ \sum _{n=0}^{\infty} \frac{3 \Omega_{w,0} T_0}{8 T^2\, n!}\left[\frac{3 H_0 t_0 (1+w)}{1+\alpha}\right]^n {_2F_1}\left[-\frac{4 \alpha}{1+\alpha},-n;1-\frac{4 \alpha}{1+\alpha};\left(\frac{T}{T_0}\right)^{\frac{1+\alpha}{2 \alpha}}\right]\,,
\label{Solution_Sec_4.4}
\end{align}
where $c_1$ is an integration constant and\, $_2F_1(a,b;c;z)$ holds again for the Gauss's hypergeometric function. We find that the first term corresponds to the Gauss-Bonnet term (since the full gravitational Lagrangian is $f = T_G g$) and hence can be removed from the Lagrangian. For completeness, the term is retained in what follows. To restrict the possible types of functions as given by \eqref{Solution_Sec_4.4}, we aim for vacuum solutions to be recovered i.e., $f(0,0) = 0$. Since $T_G \sim T^2$, we require that the hypergeometric function goes to zero when $T \rightarrow 0$. Due to the power-dependence in the argument of the hypergeometric function, we have two cases depending on the exponent $(1+\alpha)/2\alpha$. These turn out to be identical to those found in Sec. \ref{4.2}, in other words the possible set of functions are those with $-1 < \alpha < 0$ (i.e., negative exponent in the hypergeometric argument in Eq. \eqref{Solution_Sec_4.4}). To analyse the behaviour of this function, we shall examine particular limits for $T$ below. 
In \ref{AppendixB} we have studied the two limiting cases for the variable $T/T_0$ in $g(T)$ solution \eqref{Solution_Sec_4.4} in order to provide further insight about the asymptotic form of $g(T)$ for these models.

\subsection{$f(T,T_G) = -T + T_G g(T)$}
\label{4.5}
For these types of models, the Friedmann equation becomes
\begin{equation}
T - \dfrac{4T^3}{3} g_T = T_0 \Omega_{w,0} \exp \left\lbrace\dfrac{3(1+w) H_0 t_0}{1+\alpha} \left[1-\left(\frac{T}{T_0}\right)^{\frac{1+\alpha}{2\alpha}}\right] \right\rbrace.
\end{equation} 
The solution is the same as Sec. \ref{4.4} with an extra particular solution
\begin{equation}\label{eq_hubble_tegr+tg}
g_\text{part.}(T) = -\dfrac{3}{4T}.
\end{equation}
The analysis for the limits on the hypergeometric and convergence are identical to those carried out in the previous Sec. \ref{4.4}. The results of Section \ref{4.4} as summarised on Table \ref{Table2}. 


\subsection{$f(T,T_G) = -T + \mu \left(\frac{T}{T_0}\right)^\beta \left(\frac{T_G}{T_{G,0}}\right)^\gamma$}
\label{4.6}
For these types of models, the Friedmann equation becomes
\begin{align}
\label{Eq_4.6}
& T_0 \Omega_{w,0} a^{-3(1+w)} = T \nonumber \\
&+ \mu \left(\dfrac{T}{T_0}\right)^\beta \left(\dfrac{T_G}{T_{G,0}}\right)^\gamma \left[1 - 2 \beta - \gamma + \dfrac{2\alpha \beta \gamma}{\alpha+Ht}+ \dfrac{\alpha \gamma (\gamma-1)}{\left(\alpha+Ht\right)^2} \left(3\alpha-1 +4Ht\right) \right].
\end{align} 
By evaluating the expression at current times, the constant $\mu$ is found to be
\begin{equation}
\label{mu_4.6}
\mu = \dfrac{T_0 \left[\Omega_{w,0} - 1\right]}{1 - 2 \beta - \gamma + \dfrac{2\alpha \beta \gamma}{\alpha+H_0 t_0}+ \dfrac{\alpha \gamma (\gamma-1)}{\left(\alpha+H_0 t_0\right)^2} \left(3\alpha-1 +4H_0 t_0\right)},
\end{equation} 
provided that the denominator is non-zero. Assuming that $\mu\neq 0$ in \eqref{mu_4.6} holds, Eq. \eqref{Eq_4.6} becomes
\begin{align}
&\dfrac{\mu}{T_0} \left(\dfrac{t}{t_0}\right)^{2\alpha\beta} \left(\dfrac{T_G}{T_{G,0}}\right)^\gamma \left[1 - 2 \beta - \gamma + \dfrac{2\alpha \beta \gamma}{\alpha+Ht}+ \dfrac{\alpha \gamma (\gamma-1)}{\left(\alpha+Ht\right)^2} \left(3\alpha-1 +4Ht\right) \right] a^{3(1+w)} \nonumber \\
&+ \left(\dfrac{t}{t_0}\right)^{2\alpha} a^{3(1+w)} = \Omega_{w,0}.
\end{align} 
Obviously, this equation is only valid provided that the l.h.s. is time independent. For $\alpha \neq -1$, this becomes problematic due to the presence of the exponential terms which the scale factor may rise up according to Eq. \eqref{a_from_H}. An attempt for solving the time independence has been carried out, with an apparent solution to exist in the limit when $\alpha \rightarrow \infty$ whilst only applicable in certain epochs (with no clear solution for all times). However, this $\alpha$ limit results into an unphysical resolution, being that $\Omega_{\text{M},0} \rightarrow 0$, which is clearly not in agreement with observational data. Thus, no Lagrangian solution has been found for this power law model. 

The results of Section \ref{S4} are summarised in Table \ref{Table2}.
\begin{table}
\begin{tabular}{ | l | l | l |}
\hline
\hline
\multicolumn{3}{ |c| }{de Sitter solutions: $H=H_{dS} \rightarrow$ $T_{dS} = 6{H_\text{dS}}^2$ and  $T_{G_{dS}} = 24{H_\text{dS}}^4$} \\
\hline
\hline
\multirow{2}{*}{$\alpha=0$} & \multicolumn{2}{ |l| }{Any solution of the algebraic equation: $f(T_{dS},T_{G dS}) - 2T_{dS} f_{T_{dS}} - T_{G dS} f_{T_{G dS}} = 0$,} \\ 
& \multicolumn{2}{|l|}{one class of which is Eq. \eqref{Solution_deSitter}.} \\
\hline
\hline
\multicolumn{3}{ |c| }{$f(T,T_G)=g(T)+h(T_G)$} \\
\hline
\hline
\multicolumn{3}{ |c| }{\multirow{2}{*}{$g(T) = c_1 \sqrt{T}+ \Omega_{w,0} T_0 \sum\limits_{n=0}^\infty \frac{1}{n!} \left[\frac{3 H_0 t_0 (1+w)}{1+\alpha}\right]^n _2F_1\left[-\frac{\alpha}{1+\alpha},-n;\frac{1}{1+\alpha};\left(\frac{T}{T_0}\right)^{\frac{1+\alpha}{2 \alpha}}\right]$}} \\ 
\multicolumn{3}{ |c| }{}\\ \hline
\multicolumn{3}{|c|}{No exact solution for $h(T_G)$} \\
\cline{1-3}
Limits & \multicolumn{2}{ |c| }{$-1<\alpha<0$}\\ \hline
\multirow{3}{*}{$T/T_0\ll1$} & \multicolumn{2}{|c|}{\multirow{3}{*}{$g(T)=c_1 \sqrt{T}+\Omega_{w,0} T_0 e^z \sum\limits_{m=0}^\infty \dfrac{(-z)^m}{(b)_m}$;  $b \equiv  \dfrac{1}{1+\alpha}$ and $z \equiv -\dfrac{3 H_0 t_0 (1+w)}{1+\alpha}\left(\dfrac{T}{T_0}\right)^{\frac{1+\alpha}{2 \alpha}}$}} \\ 
&\multicolumn{2}{|c|}{} \\
&\multicolumn{2}{|c|}{} \\
\cline{1-3}
\multirow{3}{*}{$T/T_0\gg1$} & \multicolumn{2}{|c|}{\multirow{3}{*}{$g(T)=g(T) \approx c_1 \sqrt{T} + \Omega_{w,0} T_0 \exp \left[\frac{3 H_0 t_0 (1+w)}{\alpha+1}\right] \left[1 +\alpha \frac{3 H_0 t_0 (1+w)}{\alpha+1} \left(\frac{T}{T_0}\right)^{\frac{\alpha+1}{2 \alpha}}\right].
$}} \\ 
&\multicolumn{2}{|c|}{} \\
&\multicolumn{2}{|c|}{} \\
\hline
 & \multicolumn{1}{ |c| }{$(1+\alpha)/2\alpha>0$} & \multicolumn{1}{ |c| }{$-1<\alpha<0$} \\ \cline{2-3}
\multirow{2}{*}{$T\rightarrow 0$}  & \multicolumn{1}{ |c| }{\multirow{2}{*}{$h(T_G)\,=\,c_1\,T_G^{\frac{\alpha}{3\alpha-1}}+c_2\,T_G \qquad\qquad$ }} & \multicolumn{1}{ |c| }{\multirow{2}{*}{$h(T_G)\,=\,c_3\,T_G$}}  \\
& & \\
\hline
\hline
\multicolumn{3}{ |c| }{$f(T,T_G)=T_G g(T)$} \\
\hline
\hline
\multicolumn{3}{ |c| }{\multirow{2}{*}{$g(T) = c_1+ \sum _{n=0}^{\infty} \frac{3 \Omega_{w,0} T_0}{8 T^2 n!}\left(\frac{3 H_0 t_0 (1+w)}{1+\alpha}\right)^n {_2F_1}\left(-\frac{4 \alpha}{1+\alpha},-n;1-\frac{4 \alpha}{1+\alpha};\left(\frac{T}{T_0}\right)^{\frac{1+\alpha}{2 \alpha}}\right)$}} \\ 
\multicolumn{3}{ |c| }{}\\ \hline
Limits & \multicolumn{2}{ |c| }{$\alpha \neq (1-m)/(3+m)$ for $m=0,-1,-2, -4, -5...$}\\ \hline
\multirow{3}{*}{$T/T_0\ll1$} & \multicolumn{2}{|c|}{\multirow{3}{*}{$g(T) = c_1+\dfrac{3 \Omega_{w,0} T_0}{8 T^2}\, _1F_1\left(-\frac{4 \alpha}{1+\alpha};\frac{1-3\alpha}{1+\alpha};-\frac{3 H_0 t_0 (1+w)}{1+\alpha}\left(\frac{T}{T_0}\right)^{\frac{1+\alpha}{2\alpha}}\right).
$}} \\ 
&\multicolumn{2}{|c|}{} \\
&\multicolumn{2}{|c|}{} \\
\cline{1-3}
\multirow{3}{*}{$T/T_0\gg1$} & \multicolumn{2}{|c|}{\multirow{3}{*}{$g(T) \approx c_1 + \frac{3 \Omega_{w,0} T_0}{8 T^2} \exp \left[\frac{3 H_0 t_0 (1+w)}{1+\alpha}\right] \left[1+\frac{3 H_0 t_0 (1+w)}{1+\alpha}\dfrac{4 \alpha}{1-3 \alpha}\left(\frac{T}{T_0}\right)^{\frac{1+\alpha}{2 \alpha}}\right].$}} \\ 
&\multicolumn{2}{|c|}{} \\
&\multicolumn{2}{|c|}{} \\
\hline
\hline
\multicolumn{3}{ |c| }{$f(T,T_G)=-T+T_G g(T)$} \\
\hline
\hline
\multicolumn{3}{ |c| }{$\alpha \neq (1-m)/(3+m)$ for $m=0,-1,-2, -4, -5...$}\\ \hline
\multicolumn{3}{ |c| }{\multirow{2}{*}{$g(T) = c_1+ \sum _{n=0}^{\infty} \frac{3 \Omega_{w,0} T_0}{8 T^2 n!}\left(\frac{3 H_0 t_0 (1+w)}{1+\alpha}\right)^n {_2F_1}\left(-\frac{4 \alpha}{1+\alpha},-n;1-\frac{4 \alpha}{1+\alpha};\left(\frac{T}{T_0}\right)^{\frac{1+\alpha}{2 \alpha}}\right)-\dfrac{3}{4T}$}} \\ 
\multicolumn{3}{ |c| }{}\\ \hline
\hline
\hline
\multicolumn{3}{ |c| }{$f(T,T_G) = -T+ \mu \left(\frac{T}{T_0}\right)^\beta \left(\frac{T_G}{T_{G,0}}\right)^\gamma $} \\
\hline
\hline
\multicolumn{3}{ |c| }{No physical solution found.}\\
\hline
\hline
\end{tabular}
 \caption{Summary of the Lagrangians $f(T,T_G)$ that reproduce exponential solutions of the type $a(t)\propto \exp\left(\frac{t^{\alpha+1}}{1+\alpha}\right)$ with $\alpha\neq-1$. However, in general this type of cosmologies, except in the dS case ($\alpha=0$), leads to the presence of hypergeometric functions in the action, which may become problematic while analysing the convergence of the series or even the complexity of the function for some arguments, as well as the possible non-existence of vacuum solutions. More details in the text.}
 \label{Table2}
\end{table}

\section{Reconstruction for $\Lambda$CDM exact solution for dust matter and cosmological constant}
\label{S5}

Let us now focus on the possibility of reconstructing a gravitational Lagrangian capable of mimicking the $\Lambda$CDM analytical scale factor when both dust matter and a cosmological constant are the sole ingredients in the cosmological budget. For simplicity no spatial curvature is considered. As well known this model is described by a scale factor of the form\footnote{The relative density today in form of cosmological constant $\Omega_\Lambda$ has been substituted by $\Omega_\Lambda=1-\Omega_{{\rm M},0}$.}\cite{delaCruzDombriz:2011wn}
\begin{equation}\label{a_LCDM}
a(t) = \left(\dfrac{\Omega_{\text{M},0}}{1-\Omega_{{\rm M},0}}\right)^{1/3} \sinh^{2/3}\left(\dfrac{3 \sqrt{1-\Omega_{{\rm M},0}}}{2} H_0 t\right)\,.
\end{equation}
To lighten the notation, let us introduce the following quantities
\begin{align}
K_1 &\equiv \left(\dfrac{\Omega_{\text{M},0}}{1-\Omega_{{\rm M},0}}\right)^{1/3}, & K_2 &\equiv \dfrac{3 \sqrt{1-\Omega_{{\rm M},0}}}{2} H_0.
\end{align}
In this way, the scale factor is simply expressed as $a(t) = K_1 \sinh^{1/3} (K_2 t)$. For this scale factor, $H$, $T$ and $T_G$ are given by
\begin{align}
H &= \dfrac{2 K_2}{3} \coth\left(K_2t\right), & T &= 6H^2, & T_G &= \left(1-\Omega_{{\rm M},0}\right) T_0 T - \dfrac{T^2}{3}.\label{H_T_TG_LCDM}
\end{align}
Furthermore, the scale factor can be expressed in terms of $T$ as
\begin{equation}\label{a_T_LCDM}
a^3 = \dfrac{8 {K_1}^3 {K_2}^2}{3 T-8 {K_2}^2} = \dfrac{T_0 \Omega_{\text{M},0} }{T- T_0 \left(1-\Omega_{{\rm M},0}\right)}.
\end{equation}
Note that this expression holds for all times since $3T - 8{K_2}^2$ never cancels. Rearranging expression \eqref{a_T_LCDM} yields 
\begin{equation}\label{eq:LCDM-friedmann}
\dfrac{T}{T_0} = \Omega_{\text{M},0}a^{-3} + \left(1-\Omega_{{\rm M},0}\right)\,,
\end{equation}
i.e., the standard $\Lambda$CDM Friedmann equation with dust and cosmological constant contributions only.
Since we are interested in reconstructing $\Lambda$CDM solutions, the role of the $f(T,T_G)$ Lagrangian would be precisely to explain this cosmological evolution without invoking any dark fluid nor the addition of any cosmological constant. 


In analogy with the previous sections, in the following we shall consider different classes of $f(T,T_G)$ models. It is important to note that in absence of $T_G$ and by assuming the presence of dust matter (i.e., $w = 0$), the Friedmann equation \eqref{Friedmann1} can be solved by plugging Eq. \eqref{eq:LCDM-friedmann} where required, rendering the well-known solution 
\begin{equation}
f(T) = -T - \left(1-\Omega_{{\rm M},0}\right) T_0 + c_1 \sqrt{T},
\end{equation}
where $c_1$ is an integration constant. This is the same solution as found in \cite{Salako:2013gka} with $\left(1-\Omega_{{\rm M},0}\right) T_0 \equiv 2\Lambda$. 

\subsection{$f(T,T_G) = g(T) + h(T_G)$}
\label{5.1}
For this case, since the scale factor can be expressed in terms of $T$ only, this gives rise to two separate differential equations being
\begin{eqnarray}
&g - 2T g_T = T_0 \Omega_{w,0} \left[\dfrac{T_0 \Omega_{\text{M},0} }{T- T_0 \left(1-\Omega_{{\rm M},0}\right)}\right]^{-(1+w)}, \label{Eq_g_LCDM_5.1}\\
&h - T_G h_{T_G} + 2 \left({T_G}^2 - T_G T^2 + \dfrac{2T^4}{9}\right) h_{T_{G}T_G} = 0\label{Eq_h_LCDM_5.1}.
\end{eqnarray} 
The solution for $g$ is given by
\begin{eqnarray}
g(T) &=& c_1 \sqrt{T}+T_0 \Omega_{w,0} \left[\frac{\Omega_{\text{M},0} T_0}{T-\left(1-\Omega_{{\rm M},0}\right) T_0}\right]^{-(1+w)} \times\nonumber\\
&&\left\{\, _2F_1\left[1,w+\frac{1}{2};\frac{1}{2};\frac{T}{\left(1-\Omega_{{\rm M},0}\right) T_0}\right]\right.
\nonumber\\
&&\left.+\,\dfrac{T}{\left(1-\Omega_{{\rm M},0}\right) T_0} \, _2F_1\left[1,w+\frac{3}{2};\frac{3}{2};\frac{T}{\left(1-\Omega_{{\rm M},0}\right) T_0}\right]\right\}, 
\end{eqnarray}
where $c_1$ is an integration constant. The first term corresponds to the DGP term which is expected to appear. Before simplifying the expression above in certain limits of interest, we shall first investigate whether vacuum solutions can be recovered. For this to occur, we require $f(0,0) = 0$, i.e., either $g(0) = -h(0)$ or $g(0) = h(0) = 0$. In fact, the second case is ruled out since it turns out that
\begin{equation}
\label{g0_LCDM}
g(0) = T_0 \Omega_{w,0} \left(\frac{\Omega_{\text{M},0}}{\Omega_{{\rm M},0}-1}\right)^{-(1+w)}, 
\end{equation}
which is a non-zero constant. Interestingly, this expression yields a constraint on the possible values of $w$ since the bracketed term is negative ($\Omega_{{\rm M},0}\le 1$ according to observational data). For instance, dust provides a real value for $g(0)$ while radiation does not. 


Let us now analyse equation \eqref{Eq_h_LCDM_5.1} for $h(T_G)$. Since such equation is not fully expressed in terms of $T_G$, we can use Eq. \eqref{H_T_TG_LCDM} to obtain the relation among both scalars:
\begin{equation}
T = \dfrac{3\left(1-\Omega_{{\rm M},0}\right) T_0}{2} \left[1 \pm\sqrt{1-\dfrac{4 {T_G}}{3{\left(1-\Omega_{{\rm M},0}\right)}^2 {T_0}^2}}\right]\,.
\label{T_as_TG_LCDM}
\end{equation}
By evaluating the expression at current times and using $T = T_0$ and $T_G = {T_0}^2 \left(\frac{2}{3}-\Omega_{{\rm M},0} \right)$, yields
\begin{equation}
2 = 3\left(1-\Omega_{{\rm M},0}\right) \pm \left(1- 3\Omega_{{\rm M},0}\right)\,,
\end{equation}
leading to the negative sign in Eq. \eqref{T_as_TG_LCDM} as the only possible solution. Furthermore, solution \eqref{T_as_TG_LCDM} has to be real, meaning that
\begin{equation}
T_G \leq \dfrac{3{\left(1-\Omega_{{\rm M},0}\right)}^2 \,{T_0}^2}{4}.
\end{equation}
By the expression of $T_G$, the maximum value of $T_G$ occurs at $T = 3 \left(1-\Omega_{{\rm M},0}\right) T_0/2$ which is precisely $\dfrac{3{\left(1-\Omega_{{\rm M},0}\right)}^2 \,{T_0}^2}{4}$. Thus, expression \eqref{T_as_TG_LCDM} with the minus sign is well defined at all times. Therefore, Eq. \eqref{Eq_h_LCDM_5.1}  can be expressed solely in terms of $T_G$, as follows 
\begin{align}
\label{Eq_h_LCDM}
&0 = h - T_G h_{T_G} \nonumber \\
&+ 18{\left(1-\Omega_{{\rm M},0}\right)}^4 {T_0}^4 \left\{1-\dfrac{11 T_G}{6 {\left(1-\Omega_{{\rm M},0}\right)}^2 {T_0}^2}+\dfrac{2 {T_G}^2}{3 {\left(1-\Omega_{{\rm M},0}\right)}^4 {T_0}^4}\right.\nonumber\\
&-\left. \left[1-\dfrac{7\,T_G}{6 {\left(1-\Omega_{{\rm M},0}\right)}^2 {T_0}^2}\right] \sqrt{1-\dfrac{4 T_G}{3 {\left(1-\Omega_{{\rm M},0}\right)}^2 {T_0}^2}}\right\} h_{T_{G}T_G}.
\end{align}
%
%
whose solution turns out to be
\begin{align}
h(y) &= c_2 \left(1-y^2\right)-\frac{1}{128} c_3 (1-3 y)^{2/3} \bigg[(40y+24)\sqrt{1-y} \nonumber \\
&+ 5 \sqrt{6} \, _2F_1\left(\frac{1}{2},\frac{2}{3};\frac{5}{3};\frac{3y-1}{2}\right) (y^2-1)\bigg],
\end{align}
where $y \equiv \sqrt{1-\dfrac{4 T_G}{3{\left(1-\Omega_{{\rm M},0}\right)}^2 {T_0}^2}}$ 
where $c_{2,3}$ are integration constants. The first term corresponds to the Gauss-Bonnet which is expected to appear.  In order to investigate the vacuum solutions condition, we take the $T_G \rightarrow 0$ limit, which corresponds to $y \rightarrow 1$. It turns out that $h(T_G = 0) = 0$. This result, together with the obtained value for $g(T=0)\neq 0$ as obtained in Eq. \eqref{g0_LCDM} leads us to conclude that Lagrangians of the form studied here cannot describe vacuum solutions since the condition $f(0,0)= h(0) + g(0) = 0$ is not satisfied. 


\subsection{$f(T,T_G) = T g(T_G)$}
\label{5.2}
For $g(T_G)$ rescaling models, the Friedmann equation becomes
\begin{equation}\label{Eq_5.2}
g - T_G g_{T_G} + \dfrac{4T^2}{3} g_{T_G} - 2 \left({T_G}^2 - T_G T^2 + \dfrac{2T^4}{9}\right) g_{T_G T_G}  = -\dfrac{T_0}{T} \Omega_{w,0} \left[\dfrac{T_0 \Omega_{\text{M},0} }{T- T_0 \left(1-\Omega_{{\rm M},0}\right)}\right]^{-(1+w)}.
\end{equation} 
By using Eq. \eqref{T_as_TG_LCDM} and by introducing the variable 
 $y \equiv \sqrt{1- \dfrac{4 {T_G}}{3{\left(1-\Omega_{{\rm M},0}\right)}^2 {T_0}^2}}$\,,  
Eq. \eqref{Eq_5.2} can be recast as
\begin{align}
&2y^2 \left(y-1\right) g + (y-1)^2 \left(y^2-5 y+2\right) g_y -  2y(3 y-1) (y-1)^3 g_{yy} \nonumber \\
&= \dfrac{4y^2}{3\left(1-\Omega_{{\rm M},0}\right)}\left[\dfrac{2 \Omega_{\text{M},0}}{3\left(1-\Omega_{{\rm M},0}\right) \left(1-y\right) - 2\left(1-\Omega_{{\rm M},0}\right)}\right]^{-(1+w)}. \label{Eq_5.2-2}
\end{align}
whose solution for the associated homogeneous equation can be extracted using power series. Namely, assuming that
\begin{equation}
g_{hom}(y) \equiv \sum\limits_{n=0}^\infty a_n y^n,
\end{equation}
where $a_n$ are unknown coefficients, yields the following recurrence relation
\begin{align}
&\left(2+7n-6 n^2\right) a_n +\left(20 n^2+13 n-9\right) a_{n+1}-(n+2) (24 n+11) a_{n+2} \nonumber \\
&+3 (n+3) (4 n+5) a_{n+3}-2 (n+2) (n+4) a_{n+4} = 0,
\end{align}
which is defined for $n \geq 0$ with the conditions $a_0 = 3(a_2 - a_3)$ and $a_1 = 0$ being $a_2$ and $a_3$ the constants of integration for the homogeneous solution. For illustrative purposes we provide below the first six non-zero terms in the series 
\begin{align*}
a_0 &= 3(a_2 - a_3)\,,\; 
a_2 = c_1\,,\;
a_3 = c_2\,,\; 
a_4 = \frac{1}{16} (39 a_3-16 a_2), \\
a_5 &= \frac{1}{40} (211 a_3-112 a_2)\,,\;
a_6 = \frac{1}{384} (4403 a_3-2544 a_2),
\end{align*}
where we have defined $a_2$ and $a_3$ in terms of $c_1$ and $c_2$ respectively to denote them as the integration constants. Thus, the homogeneous solution is given by
\begin{align}
g_{hom}(y) &= c_1 \left(3 + y^2 - y^4 - \dfrac{14}{5}y^5 - \dfrac{53}{8}y^6 \dots \right) \nonumber \\
&+ c_2 \left(-3 + y^3 +\dfrac{39}{16}y^4 + \dfrac{211}{40}y^5 + \dfrac{4403}{384}y^6 \dots\right).
\end{align}
which does not correspond to any well-known series in terms of transcendental functions. In principle, since the homogeneous solution has been obtained, the particular solution can be found using the well-known Green's function and Wronskian method \cite{kythe2001green}. However, since the homogeneous solutions are expressed in terms of power series,  the particular solution would require numerical methods in addition to boundary conditions, which are unknown. 


\subsection{$f(T,T_G) = T_G g(T)$}
\label{5.3}
For this model, the Friedmann equation reduces to
\begin{equation}\label{Eq_5.3}
- \dfrac{4 T^3}{3}g_T = T_0 \Omega_{w,0} \left[\dfrac{T_0 \Omega_{\text{M},0}}{T- T_0 \left(1-\Omega_{{\rm M},0}\right)}\right]^{-(1+w)},
\end{equation} 
being the solution given by
\begin{eqnarray}
\label{Sol_5.3}
g(T) &=& c_1+\frac{3 \Omega_{w,0} T_0}{8 T^2}\left[\frac{\Omega_{\text{M},0} T_0}{T-\left(1-\Omega_{{\rm M},0}\right) T_0}\right]^{-(1+w)}\nonumber\\
&& \times \left\{1+\dfrac{1+w}{1-w} \, _2F_1\left[1,2;2-w;\frac{\left(1-\Omega_{{\rm M},0}\right) T_0}{T}\right]\right\}, 
\end{eqnarray}
where $c_1$ is an integration constant, which corresponds to the Gauss-Bonnet term in the Lagrangian. Note that due to the presence of the hypergeometric function, the solution is not valid for $w = n$, where $n \in \mathbb{Z}^{+}$, as the solution becomes singular. 
%
%
For instance, for $w = 1$, the solution of \eqref{Eq_5.3} is given by
\begin{equation}\label{Sol_5.3_w=1}
g(T) = c_1-\frac{3 \Omega_{w,0}}{4 {\Omega_{\text{M},0}}^2 T_0}\left[-\frac{{\left(1-\Omega_{{\rm M},0}\right)}^2 {T_0}^2}{2 T^2}+\frac{2 \left(1-\Omega_{{\rm M},0}\right) T_0}{T}+\ln \left(\dfrac{T}{T_0}\right)\right].
\end{equation}
However, this solution this does not generate vacuum solutions since the Lagrangian diverges at $T\rightarrow0$ as 
\begin{equation}
f(0) = \frac{\Omega_{w,0}{\left(1-\Omega_{{\rm M},0}\right)}^2 {T_0}}{8 {\Omega_{\text{M},0}}^2}\left[3 \left(1-\Omega_{{\rm M},0}\right) \dfrac{T_0}{T}-13\right]\bigg|_{T \rightarrow 0}.
\end{equation}
This seems to be the case for all other positive integer values. Indeed, by assuming that $w = n$ for integer $n \geq 1$, then Eq. \eqref{Eq_5.3} becomes
\begin{equation}
- \dfrac{4 T^3}{3}g_T = 
T_0 \Omega_{w,0} \left[\dfrac{T- T_0 \left(1-\Omega_{{\rm M},0}\right)}{T_0 \Omega_{\text{M},0}}\right]^{1+n}.
\end{equation} 
Evaluating the function at $T = 0$ yields
\begin{equation}
T^3 g_T|_{T = 0} = \dfrac{3}{4} T_0 \Omega_{w,0} (-1)^{n} \left(\dfrac{1-\Omega_{{\rm M},0}}{\Omega_{\text{M},0}}\right)^{1+n}.
\end{equation}
This suggests that $T^3 g_T \sim T^2 g \sim T_G g = f$ at $T = 0$ is non-zero, hence leading to no vacuum solutions. \newline

For the hypergeometric solution in Eq. \eqref{Sol_5.3}, this also suffers from recovering vacuum solutions. In this case the Lagrangian, reduces to
\begin{eqnarray}
f(0) &=& \frac{3 \Omega_{w,0} \left(1-\Omega_{{\rm M},0}\right)  {T_0}^2}{8 T}\left[\frac{\Omega_{\text{M},0} T_0}{T-\left(1-\Omega_{{\rm M},0}\right) T_0}\right]^{-(1+w)} \nonumber\\
&&-  \frac{\Omega_{w,0} {T_0}}{8}\left(-\frac{\Omega_{\text{M},0}}{1-\Omega_{{\rm M},0}}\right)^{-(1+w)} (4+3w)\bigg|_{T\rightarrow 0},
\end{eqnarray}
which is undefined for every $w \neq n$, where $n \in \mathbb{Z}^{+}$ (we are excluding these values since Eq. \eqref{Sol_5.3} is not defined at these instances), due to the first term which tends to infinity. Thus, these Lagrangian models in Sec. \ref{5.3} cannot describe the sought $\Lambda$CDM model solution if vacuum solutions are expected to be recovered.

\subsection{$f(T,T_G) = -T + T_G g(T)$}
\label{5.4}
For this model, the Friedmann equation reduces to
\begin{equation}
T - \dfrac{4 T^3}{3}g_T = T_0 \Omega_{w,0} \left[\dfrac{T_0 \Omega_{\text{M},0}}{T- T_0 \left(1-\Omega_{{\rm M},0}\right)}\right]^{-(1+w)},
\end{equation} 
The solution is the same as the previous section as found in Eq. \eqref{Sol_5.3} for $w \neq n$, $n \in \mathbb{Z}^+$ and the special cases for $w = n$, $n \in \mathbb{Z}^+$, which are solved separately (for instance, the solution as given for $w = 1$ in Eq. \eqref{Sol_5.3_w=1}), but one must add an extra particular solution
\begin{equation}
g_\text{part.}(T) = -\dfrac{3}{4T}.
\end{equation}
The resulting conclusions about vacuum solutions are identical to those found in the previous section, Section \ref{5.3}, i.e. we are led to conclude that the obtained gravitational Lagrangians cannot host vacuum solutions due to the divergence as $T \rightarrow 0$ for these solutions.

\subsection{$f(T,T_G) = -T + \mu\left(\frac{T}{T_0}\right)^\beta\left(\frac{T_G}{T_{G,0}}\right)^\gamma$}
\label{5.5}
For power-law $f(T,T_G)$ models in both $T$ and $T_G$, the Friedmann equation becomes
\begin{align}\label{Eq_5.5}
&T + \mu\left(\dfrac{T}{T_0}\right)^{\beta+\gamma}\left[\dfrac{3\left(1-\Omega_{{\rm M},0}\right) T_0 - T}{  \left(2 - 3 \Omega_{{\rm M},0}\right)T_0 }\right]^\gamma\bigg[(\gamma-1)(2\gamma+2\beta-1) \nonumber \\
&-\dfrac{4\beta \gamma +6\gamma(\gamma-1)}{3\left(1-\Omega_{{\rm M},0}\right) T_0 - T}\,T+\frac{4\gamma(\gamma-1) T^2}{(3 \left(1-\Omega_{{\rm M},0}\right) T_0-T)^2}\bigg] = T_0 \Omega_{w,0} \left[\dfrac{T_0 \Omega_{\text{M},0}}{T- T_0 \left(1-\Omega_{{\rm M},0}\right)}\right]^{-(1+w)},
\end{align}
for some constants $\mu$, $\beta$ and $\gamma$. Note that in order for vacuum solutions to be obtained, we require
\begin{equation}
\beta + \gamma > 0.
\end{equation}
The value for $\mu$ can be found by evaluating Eq. \eqref{Eq_5.5}
 at current times, yielding
\begin{equation}
\mu = \dfrac{T_0 \left(\Omega_{w,0}-1\right)}{(\gamma-1)(2\gamma+2\beta-1)-\dfrac{4\beta \gamma +6\gamma(\gamma-1)}{2 - 3\Omega_{{\rm M},0}}+\dfrac{4\gamma(\gamma-1)}{(2- 3\Omega_{{\rm M},0})^2}} \equiv \dfrac{T_0}{\nu} \left(\Omega_{w,0}-1\right),
\end{equation}
provided that the denominator $\nu$ is non-zero. Rearranging Eq. \eqref{Eq_5.5}, it yields
\begin{align}
\label{Eq_larga}
&\dfrac{T}{T_0}\left[\dfrac{T_0 \Omega_{\text{M},0}}{T- T_0 \left(1-\Omega_{{\rm M},0}\right)}\right]^{1+w}\nonumber\\
& + \dfrac{\left(\Omega_{w,0}-1\right)}{\nu}\left(\dfrac{T}{T_0}\right)^{\beta+\gamma}\left[\dfrac{3\left(1-\Omega_{{\rm M},0}\right) T_0 - T}{\left(2-3\Omega_{{\rm M},0}\right) T_0}\right]^\gamma \left[\dfrac{T_0 \Omega_{\text{M},0}}{T- T_0 \left(1-\Omega_{{\rm M},0}\right)}\right]^{1+w}\nonumber \\
&\times \bigg[(\gamma-1)(2\gamma+2\beta-1)-\dfrac{4\beta \gamma +6\gamma(\gamma-1)}{3\left(1-\Omega_{{\rm M},0}\right) T_0 - T}\,T+\frac{4\gamma(\gamma-1) T^2}{(3 \left(1-\Omega_{{\rm M},0}\right) T_0-T)^2}\bigg] = \Omega_{w,0}.
\end{align}
Since the r.h.s in the expression above is constant, then the $T$ dependent quantities on the l.h.s. must vanish. Note that for $w \neq 0$, the first term and the $\Omega_{w,0}$ independent contribution resulting from the second term on the l.h.s cannot contribute to the r.h.s and it must cancel. This results into the following condition
\begin{align}
&\left[\dfrac{T_0 \Omega_{\text{M},0}}{T- T_0 \left(1-\Omega_{{\rm M},0}\right)}\right]^{1+w} \bigg\lbrace \nu - \left(\dfrac{T}{T_0}\right)^{\beta+\gamma-1}\left[\dfrac{3\left(1-\Omega_{{\rm M},0}\right) T_0 - T}{3\left(1-\Omega_{{\rm M},0}\right) T_0 - T_0}\right]^\gamma \nonumber \\
&\times \bigg[(\gamma-1)(2\gamma+2\beta-1)-\dfrac{4\beta \gamma +6\gamma(\gamma-1)}{3\left(1-\Omega_{{\rm M},0}\right) T_0 - T}T+\frac{4\gamma(\gamma-1) T^2}{(3 \left(1-\Omega_{{\rm M},0}\right) T_0-T)^2}\bigg]\bigg\rbrace = 0.
\end{align}
The common square-bracket factor in the first line cannot be zero and hence the curly bracketed term must be zero. Since $\nu$ is constant, the other terms must contribute to a constant. This is only possible when only one of the terms in the square bracket contributes, otherwise other remaining $T$ terms would remain resulting the equation to still be time dependent. Another way to look at the problem would be algebraically; since the torsional terms result into a constant, this is identical to finding the variable whose derivative with $T$ is zero and then solve the resulting expression (as the constant expression would be its resulting integral, with $\nu$ being the integration constant). When the solutions are found, one then only needs to verify that $\nu \neq 0$. With this approach, the resulting equation to satisfy would be
\begin{align}
&\frac{{T_0}}{T^2 [3 (1-\Omega_{\text{M},0}) {T_0}-T]^3} \left[\frac{3 (1-\Omega_{\text{M},0}) {T_0}-T}{3 (1-\Omega_{\text{M},0}) {T_0}-T_0}\right]^{\gamma } \left(\frac{T}{{T_0}}\right)^{\beta +\gamma} \nonumber \\
&\times \left(X_1+ X_2 T + X_3 T^2+ X_4 T^3\right) = 0,
\end{align}
where 
\begin{align}
&X_1 \equiv 27 (\Omega_{\text{M},0}-1)^3 {T_0}^3(\gamma -1) (\beta +\gamma -1) (2 \beta +2 \gamma -1), \\
&X_2 \equiv 9 (\Omega_{\text{M},0}-1)^2 {T_0}^2 \left\lbrace 2 \beta ^2 (5 \gamma -3)+\beta  \left[\gamma  (24 \gamma -29)+9\right]+(\gamma -1) \left[2 \gamma  (7 \gamma -5)+3\right]\right\rbrace, \\
&X_3 \equiv 3 (\Omega_{\text{M},0}-1) {T_0}\left\lbrace 2 \beta ^2 (7 \gamma -3)+\beta  \left[\gamma  (44 \gamma -45)+9\right]+(\gamma -1) \left[\gamma  (32 \gamma -13)+3\right]\right\rbrace, \\
&X_4 \equiv (\beta +2 \gamma -1) \left[\beta  (6 \gamma -2)+(\gamma -1) (12 \gamma -1)\right].
\end{align}
The only possible solution for all times reduces each coefficient to zero, i.e. $X_1 = X_2 = X_3 = X_4 = 0$. This results into a system of equations, whose solutions are $\left\{\beta = \frac{1}{2},\gamma = 0\right\},\{\beta = 1,\gamma = 0\},\{\beta = -1,\gamma = 1\},\{\beta = 0,\gamma = 1\}$. The cases $\left\{\beta = \frac{1}{2},\gamma = 0\right\}$ and $\{\beta = 0,\gamma = 1\}$ correspond to the DGP and Gauss-Bonnet terms respectively, which rightfully do not obey the condition $\nu \neq 0$ since they do not contribute to the Friedmann equation. Nevertheless, it is straightforward that no vacuum solutions are allowed under such conditions.

Let us consider the case when $w = 0$. In this case, the Friedmann equation reduces to
\begin{align}
&\left(\dfrac{T}{T_0}\right)^{\beta+\gamma}\left[\dfrac{3\left(1-\Omega_{{\rm M},0}\right) T_0 - T}{\left(2-3\Omega_{{\rm M},0}\right) T_0 }\right]^\gamma \nonumber \\
&\times \left\{(\gamma-1)(2\gamma+2\beta-1)-\dfrac{4\beta \gamma +6\gamma(\gamma-1)}{3\left(1-\Omega_{{\rm M},0}\right) T_0 - T}T+\frac{4\gamma(\gamma-1) T^2}{[3 \left(1-\Omega_{{\rm M},0}\right) T_0-T]^2}\right\}= \nu\,,
\end{align}
where in order for the l.h.s. to be constant, the possible values for $\beta$ and $\gamma$ are either $\{\gamma = 1, \beta = -2\}$ or $\{\gamma = 0, \beta = 0\}$. However, in both instances, the condition for vacuum solutions is not satisfied. 
The summary of Lagrangians that reproduce $\Lambda$CDM solution are summarised in Table \ref{Table3}.
\begin{table}
\begin{tabular}{ | l | l | l |}
\hline
\hline
\multicolumn{3}{ |c| }{$f(T,T_G)=g(T)+h(T_G)$} \\
\hline
\hline
\multicolumn{3}{ |c| }{\multirow{2}{*}{$g(T)= c_1 \sqrt{T}+T_0 \Omega_{w,0} \left(\frac{\Omega_{\text{M},0} T_0}{T-\Omega_\Lambda T_0}\right)^{-(1+w)} \bigg[\, _2F_1\left(1,w+\frac{1}{2};\frac{1}{2};\frac{T}{\Omega_\Lambda T_0}\right)+\dfrac{T}{\Omega_\Lambda T_0} \, _2F_1\left(1,w+\frac{3}{2};\frac{3}{2};\frac{T}{\Omega_\Lambda T_0}\right) \bigg]$}} \\ 
\multicolumn{3}{ |c| }{}\\ \hline
\multicolumn{3}{ |c| }{\multirow{2}{*}{$h(y)= c_1 \left(1-y^2\right)-\frac{1}{128} c_2 (1-3 y)^{2/3} \bigg[(40y+24)\sqrt{1-y}+ 5 \sqrt{6} \, _2F_1\left(\frac{1}{2},\frac{2}{3};\frac{5}{3};\frac{3y-1}{2}\right) (y^2-1)\bigg],$}} \\ 
\multicolumn{3}{ |c| }{}\\
\multicolumn{3}{ |c| }{{\multirow{2}{*}{where $y \equiv \sqrt{1-\dfrac{4 T_G}{3{\Omega_\Lambda}^2 {T_0}^2}}$}}}\\ 
\multicolumn{3}{ |c| }{}\\
\hline
\hline
\multicolumn{3}{c}{} \\
\hline
\hline
\multicolumn{3}{ |c| }{$f(T,T_G)=Tg(T_G)$} \\
\hline
\hline
\multicolumn{3}{ |c| }{\multirow{2}{*}{$g(T_G) = \sum\limits_{n=0}^\infty a_n y^n$ where $y \equiv \sqrt{1-\dfrac{4 T_G}{3{\Omega_\Lambda}^2 {T_0}^2}}$\,, and}} \\ 
\multicolumn{3}{ |c| }{}\\ 
\multicolumn{3}{ |l| }{\multirow{1}{*}{$\left(2+7n-6 n^2\right) a_n +\left(20 n^2+13 n-9\right) a_{n+1}$}} \\ 
\multicolumn{3}{ |l| }{\multirow{1}{*}{$-(n+2) (24 n+11) a_{n+2}+3 (n+3) (4 n+5) a_{n+3}-2 (n+2) (n+4) a_{n+4} = 0$}} \\ 
\hline
\hline
\multicolumn{3}{c}{} \\
\hline
\hline
\multicolumn{3}{ |c| }{$f(T,T_G)=T_Gg(T)$} \\
\hline
\hline
\multicolumn{3}{ |c| }{\multirow{2}{*}{$g(T) = c_1+\frac{3 \Omega_{w,0} T_0}{8 T^2}\left(\frac{\Omega_{\text{M},0} T_0}{T-\Omega_\Lambda T_0}\right)^{-(1+w)} \left[1+\dfrac{1+w}{1-w} \, _2F_1\left(1,2;2-w;\frac{\Omega_\Lambda T_0}{T}\right)\right]$, 
}} \\ 
\multicolumn{3}{ |c| }{}\\ 
\hline
\hline
\multicolumn{3}{c}{} \\
\hline
\hline
\multicolumn{3}{ |c| }{$f(T,T_G)=-T+T_G g(T)$} \\
\hline
\hline
\multicolumn{3}{ |c| }{\multirow{2}{*}{$g(T) = c_1+\frac{3 \Omega_{w,0} T_0}{8 T^2}\left(\frac{\Omega_{\text{M},0} T_0}{T-\Omega_\Lambda T_0}\right)^{-(1+w)} \left[1+\dfrac{1+w}{1-w} \, _2F_1\left(1,2;2-w;\frac{\Omega_\Lambda T_0}{T}\right)\right]-\dfrac{3}{4T}$, 
}} \\ 
\multicolumn{3}{ |c| }{}\\ 
\hline
\hline
\multicolumn{3}{c}{} \\
\hline
\hline
\multicolumn{3}{ |c| }{$f(T,T_G) = -T+ \mu \left(\frac{T}{T_0}\right)^\beta \left(\frac{T_G}{T_{G,0}}\right)^\gamma $} \\
\hline
\hline
\multicolumn{3}{ |c| }{No vacuum solution found}\\ \cline{1-3}
$w=0$ &\multicolumn{2}{|c|}{$\{\gamma, \beta\}=\{0,1\}$} \\  \hline
$w\neq 0$ &\multicolumn{2}{|c|}{$\{\gamma, \beta\}=\{1,-1\},\{1,-2\}, \{0,0\}$} \\ \hline
\hline
\hline
\end{tabular}
 \caption{Summary of the Lagrangians $f(T,T_G)$ that reproduce exact $\Lambda$CDM model. Here hypergeometric functions become also common in the action. The unique case with no hypergeometric functions, the last one, provides a physical action but with no vacuum solutions, an expected result when comparing to the same case in GR, which do not have vacuum solutions either.}
 \label{Table3}
\end{table}

\section{Conclusions}
\label{S8}
In this work we have expanded on previous studies for some extensions of Teleparallel Theories of Gravity by including arbitrary functions of the  torsion scalar and of a boundary term analogous to the Gauss-Bonnet invariant in 3+1 dimensions. 
Thus we have provided a careful description of gravitational models capable of reconstructing well-known and paradigmatic cosmological solutions. Essentially, we have explored different ways of reconstructing the corresponding Lagrangian by using either the cosmological scale factor or the Hubble parameter, whenever one of these quantities is given in terms of analytical functions. Following this, we have considered some generic gravitational Lagrangians as well as developed some techniques to obtain the exact form of such Lagrangians within this class of modified Teleparallel gravities. Furthermore, we have explored whether the obtained Lagrangians can be considered as physically relevant by analysing some of their properties. For instance, the existence of null torsion solutions as vacuum solutions, which guarantees that such Lagrangians will indeed host both Minkowski and Schwarzschild as cosmological solutions; a reasonable requirement for viable theories of gravity.\\

Hence, firstly we have considered power-law solutions in cosmology, as such a type of evolution is usually reproduced with perfect fluids with constant equation of state, such as dust or radiation. In doing so, 
several forms of the Lagrangians have been assumed. It turns out that several classes of Lagrangians can reproduce these types of solutions. However, as shown in Table \ref{Table1}, some of these Lagrangians are capable of reproducing power-law solutions in cosmology, though they do not lead to vacuum solutions, unless some restrictions on the free parameters are put in place. Other cases do not provide much information about the Lagrangian, such as the one  summarised at the bottom panel of Table \ref{Table1}. \\

In addition, other interesting cosmologies have been explored throughout the course of this paper. In particular, we have considered exponential-type solutions, which include de Sitter cosmologies as a specific case as well as the so-called {\it Little Rip} scenario. Besides the de Sitter solution, which transforms the differential field equations into an algebraic one, the other scenarios lead to complicated solutions for the gravitational actions which usually involve hypergeometric functions, as shown in Table \ref{Table2}. Nevertheless, note that this class of Lagrangians is very common in modified gravities, for instance in $f(R)$ gravity \cite{delaCruzDombriz:2006fj}. Moreover, by assuming some asymptotic limits, the approximate Lagrangian may become simpler. Up to this point, the vacuum solutions have been explored. These are contained in Table \ref{Table2}, where some of these solutions are also new solutions in $f(T)$ gravity without the Gauss-Bonnet modification (for certain free parameter choices). This is not the case for all the solutions in Table \ref{Table2}.\\

Finally, an exact $\Lambda$CDM cosmology has been considered in Section \ref{S5}. Note that the gravitational actions found here and summarised in Table \ref{Table3} are the only ones in the paper which lead to a $\Lambda$CDM cosmological evolution, while previous actions give other types of solutions, which may provide a realistic evolution after including some other elements. Then, our results from Section \ref{S5} conclude that besides the usual Teleparallel Equivalent of General Relativity with a cosmological constant, other non-trivial $f(T,T_G)$ gravitational actions without a cosmological constant can indeed mimic the Concordance Model cosmological evolution, with both dust and a cosmological constant. This is one of the most relevant results of this research.  
As in the previous case, most of these Lagrangians become either hypergeometric functions or power series, which makes their analysis difficult, as shown in Table \ref{Table3}. Nevertheless, by taking some limits the obtained Lagrangians may be further simplified as in the previous case. Moreover, the last case at the bottom of Table \ref{Table3} illustrates the existence of solutions for a simpler Lagrangian, similarly as in usual Gauss-Bonnet gravities \cite{Elizalde:2010jx}, but with second order field equations, which can prevent the existence of ghosts contrary to the general case of Gauss-Bonnet gravity \cite{Calcagni:2006ye}. Nevertheless, for most of the Lagrangians summarised in Table \ref{Table3},  apart from the cases reducing to the standard Teleparallel Equivalent of General Relativity, vacuum solutions do not exist for the remaining cases.\\

Hence, we have shown that a wide range of $f(T,T_G)$ actions can be reconstructed in order to provide paradigmatic expansion solutions. We have thus provided here an extensive analysis under which Lagrangians may be considered viable in cosmological scales. Following this, we may conclude that such extensions of Teleparallel gravity deserve further analysis, since 
a wide range of suitable Lagrangians with well-behaved expansion history have been found. Ergo, the analysis of bouncing cosmologies and the possibility of mimicking multi-fluid scenarios are merely two straightforward generalisations to be grappled with, which may shed further light on the viability of classes of $f(T,T_G)$ Teleparallel gravity theories.

%

%

%
\ack
This article is based upon work from CANTATA COST (European Cooperation in Science and Technology) action CA15117,  EU Framework Programme Horizon 2020.
This work is also supported by CSIC I-LINK1019 Project, Spanish Ministry of Economy and Science (AdlCD and DSG).
%
%
%
AdlCD acknowledges financial support from Consolider-Ingenio
MULTIDARK CSD2009-00064, FPA2014- 53375-C2-1-P  Spanish Ministry of Economy and Science, 
FIS2016-78859-P, European Regional Development Fund and Spanish Research Agency (AEI), University of Cape Town Launching Grants Programme and National Research Foundation grant 99077 2016-2018, Ref. No. CSUR150628121624 and NRF Incentive Funding for Rated Researchers (IPRR), Ref. No. IFR170131220846.
DSG is funded by the Juan de la Cierva program (Spain) No.~IJCI-2014-21733 and by MINECO (Spain), project FIS2013-44881.
%
%

\appendix

\section{Asymptotic behaviour for the Hubble power law scenario additive model (Section \ref{4.2})}
\label{AppendixA}

In this appendix section, limits for the Gauss's Hypergeometric function are investigated, in particular for the solution given by Eq. \eqref{solution_g_Add_Hubble} found in Section \ref{4.2} , the additive model for Hubble power-law models. Due to the form of the last argument in the hypergeometric function, it is interesting to investigate the behaviour in $T/T_0 \ll1$ and $T/T_0 \gg 1$ limits, which simplifies the solution for these particular epochs.  Following the discussions of vacuum solutions in Section \ref{4.2}, the following analysis is carried out for $-1 < \alpha < 0$.

\begin{itemize}
\item {\bf $T/T_0 \ll 1$}: 
Since  $\dfrac{1+\alpha}{2\alpha} < 0$ the argument in the hypergeometric solution as given by Eq. \eqref{solution_g_Add_Hubble} may become very large in this limit. This poses a problem since the hypergeometric function is not specifically defined in these domains and has to be analytically continued. For this reason, we resort in solving the Eq. \eqref{eq:powerH-add-gODE}  for $g$ within this constraint. Given that $T/T_0 \ll 1$ , then Eq. \eqref{eq:powerH-add-gODE} approximately becomes
\begin{equation}
g - 2T g_T \approx T_0 \Omega_{w,0} \exp \left\lbrace -\dfrac{3(1+w) H_0 t_0}{1+\alpha} \left(\frac{T}{T_0}\right)^{\frac{1+\alpha}{2\alpha}} \right\rbrace.
\end{equation}
Expanding the exponential into a power series and solving the previous equation yields
\begin{equation}\label{sol_g_TT0_smaller_1}
g(T) = c_1 \sqrt{T}-\alpha \Omega_{w,0} T_0\sum\limits_{n=0}^{\infty}\frac{1}{n! \left[\alpha (n-1)+n\right]}\left[-\frac{3 H_0 t_0 (1+w)}{1+\alpha}\left(\frac{T}{T_0}\right)^{\frac{1+\alpha}{2 \alpha}}\right]^n.
\end{equation}
Note that for the values of $\alpha$ considered, the denominators above do not vanish. This particular series can be represented by the confluent Kummer hypergeometric function of the first kind $_1F_1(a;b;z)$, and hence the solution \eqref{sol_g_TT0_smaller_1} yields
\begin{equation}\label{sol_g_TT0_smaller_1_simplified}
g(T) = c_1 \sqrt{T}+\Omega_{w,0} T_0 \,\, _1F_1\left[-\dfrac{\alpha}{1+\alpha};\dfrac{1}{1+\alpha};-\frac{3 H_0 t_0 (1+w)}{1+\alpha}\left(\frac{T}{T_0}\right)^{\frac{1+\alpha}{2 \alpha}}\right].
\end{equation}
The series converges provided that $b \equiv 1/(1+\alpha)$ is a non-negative (including zero) integer. 
Since $-1 < \alpha < 0$, $b$ is always positive and hence the series is convergent (as expected from the series expansion). Note that the hypergeometric function above can be further simplified as follows. The Kummer function has the property that
\begin{equation}
_1F_1(a;b;y) = {\rm e}^y \, _1F_1(b-a;b;-y),
\end{equation}
which in this case simplifies to
\begin{equation}
_1F_1(a;b;y) = {\rm e}
^y \, _1F_1(1;b;-y).
\end{equation}
From the definition of the Kummer function, we have
\begin{equation}
_1F_1(1;b;-y) = \sum\limits_{m=0}^\infty \dfrac{(1)_m}{(b)_m}\dfrac{(-y)^m}{m!} = \sum\limits_{m=0}^\infty \dfrac{(-y)^m}{(b)_m},
\end{equation}
where we have used the fact that $(1)_m = m!$ and $(b)_m \equiv \Gamma(b+m)/\Gamma(b)$ holds for the Pochhammer symbol, with $\Gamma(x)$ being the Gamma function which is defined as
\begin{equation}
\Gamma(x) \equiv \int\limits_0^\infty t^{x-1}e^{-t} \ dt.
\end{equation} 
Thus, the solution \eqref{sol_g_TT0_smaller_1_simplified} becomes
\begin{equation}\label{eq_hubble_add_g_TT0_less}
g(T) = c_1 \sqrt{T}+\Omega_{w,0} T_0\, {\rm e}^y \sum\limits_{m=0}^\infty \dfrac{(-y)^m}{(b)_m},
\end{equation}
where in our case,  $b \equiv  \dfrac{1}{1+\alpha}$ and $y \equiv -\dfrac{3 H_0 t_0 (1+w)}{1+\alpha}\left(\dfrac{T}{T_0}\right)^{\frac{1+\alpha}{2 \alpha}}$.

\item {\bf $T/T_0 \gg 1$}:
For these epochs, the argument in the hypergeometric function of Eq. \eqref{solution_g_Add_Hubble} may consequently become very small. In this case, the hypergeometric function $_2F_1(a,b;c;y)$ can be represented as an infinite power-series
\begin{equation}
_2F_1(a,b;c;y) = \sum\limits_{m=0}^{\infty} \dfrac{(a)_m (b)_m}{(c)_m}\dfrac{y^m}{m!} = 1+\dfrac{ab}{c}\dfrac{y}{1!}+\dfrac{a(a+1)b(b+1)}{c(c+1)}\dfrac{y^2}{2!}+...,
\end{equation}
Given that the argument $y\equiv T/T_0$ is small, the leading order terms in the series would be the first two terms. Thus, the solution \eqref{solution_g_Add_Hubble} approximates to	
\begin{equation}
g(T) \approx c_1 \sqrt{T} + \Omega_{w,0} T_0 \sum\limits_{n=0}^\infty \frac{1}{n!} \left[\frac{3 H_0 t_0 (1+w)}{\alpha+1}\right]^n \left[1+\alpha n\left(\frac{T}{T_0}\right)^{\frac{\alpha+1}{2 \alpha}}\right].
\end{equation}
By the use of the series expansions of $e^x$ and $xe^x$, the solution above further simplifies to a much simpler form
\begin{equation}\label{eq_hubble_add_g_TT0_more}
g(T) \approx c_1 \sqrt{T} + \Omega_{w,0} T_0 \exp \left[\frac{3 H_0 t_0 (1+w)}{\alpha+1}\right] \left[1 +\alpha \frac{3 H_0 t_0 (1+w)}{\alpha+1} \left(\frac{T}{T_0}\right)^{\frac{\alpha+1}{2 \alpha}}\right].
\end{equation}
If the ratio $T/T_0$ is sufficiently small, then the solution above approximates to a constant once the DGP term has been substracted.\newline
\end{itemize}

\section{Asymptotic behaviour for the Hubble power law scenario $T_G$ rescaling model (Section \ref{4.4})}
\label{AppendixB}

Similar to \ref{AppendixA}, the $T/T_0 \ll 1$ and $T/T_0 \gg 1$ limits are again investigated for the solution found in Section \ref{4.4} Eq. \eqref{Solution_Sec_4.4}, which is expressed in terms of Gauss's Hypergeometric function. In what follows, the values of $-1 < \alpha < 0$ are considered following the discussions given in Section \ref{4.4}.

\begin{itemize}
\item {$T/T_0 \ll 1$}: In this scenario, the argument of the hypergeometric function in Eq. \eqref{Solution_Sec_4.4} is large\footnote{Remember that the exponent $\frac{1+\alpha}{2 \alpha}$ is negative.}. Thus, the hypergeometric function cannot be expanded into a power-series since in these domains, the function needs to be analytically continued. Instead, we investigate the original Eq. \eqref{Eq_Sec_4.4} and consider this limit. Thus Eq. \eqref{Eq_Sec_4.4} approximately becomes
\begin{equation}
- \dfrac{4T^3}{3} g_T \approx T_0 \Omega_{w,0} \exp \left[-\dfrac{3(1+w) H_0 t_0}{1+\alpha} \left(\frac{T}{T_0}\right)^{\frac{1+\alpha}{2\alpha}} \right].
\end{equation}
Expanding the exponential as a power-series and solving yields
\begin{equation}
g(T) \approx c_1-\dfrac{3\alpha \Omega_{w,0} T_0}{2 T^2}\sum\limits_{n=0}^{\infty} \frac{1}{n! [\alpha (n-4)+n]}\left[-\frac{3 H_0 t_0 (1+w)}{1+\alpha}\left(\frac{T}{T_0}\right)^{\frac{1+\alpha}{2 \alpha}}\right]^n\,.
\end{equation}
Note that the denominator is always non-zero for the $\alpha$ range provided, therefore the sum above is always well-defined. This series can be expressed in terms of the Kummer confluent hypergeometric function rendering
\begin{equation}\label{Sol_g_T_4.4.1}
g(T) \approx c_1+\dfrac{3 \Omega_{w,0} T_0}{8 T^2}\, _1F_1\left[-\frac{4 \alpha}{1+\alpha};\frac{1-3\alpha}{1+\alpha};-\frac{3 H_0 t_0 (1+w)}{1+\alpha}\left(\frac{T}{T_0}\right)^{\frac{1+\alpha}{2\alpha}}\right]\,.
\end{equation}
The series converges for $b \equiv (1-3\alpha)/(1+\alpha)$ being a non-negative (including zero) integer. For the $\alpha$ range of interest, $b$ is always positive so the series is always convergent as expected from the performed analysis. Note that the hypergeometric function can be simplified as was done in \ref{AppendixA}. The Kummer function has the property that
\begin{equation}
_1F_1(a;b;z) = {\rm e}^z \, _1F_1(b-a;b;-z),
\end{equation}
which in this case simplifies to
\begin{equation}
_1F_1(a;b;z) = {\rm e}^z \, _1F_1(1;b;-z) = {\rm e}^z \sum\limits_{m=0}^\infty \dfrac{(-z)^m}{(b)_m}.
\end{equation}
Therefore, the solution \eqref{Sol_g_T_4.4.1} for $g(T)$ can also be given as
\begin{align}\label{eq_hubble_tgrescale_g_TT0_less}
g(T) = c_1 \sqrt{T}+\Omega_{w,0} T_0 \, {\rm e}^z \sum\limits_{m=0}^\infty \dfrac{(-z)^m}{(b)_m},
\end{align}
where in our case $b \equiv \dfrac{1-3\alpha}{1+\alpha}$ and $z \equiv-\dfrac{3 H_0 t_0 (1+w)}{1+\alpha}\left(\dfrac{T}{T_0}\right)^{\frac{1+\alpha}{2\alpha}}$.

\item{$T/T_0 \gg 1$}: For this limit, the argument  of the hypergeometric function in Eq. \eqref{Solution_Sec_4.4} may be small as powered to a negative exponent. Therefore, the hypergeometric function can be expressed as a power series. However, since the argument is small, we could assume that only the first two terms in the series would contribute to the expansion. Thus, the solution of Eq. \eqref{Eq_Sec_4.4}  approximates to
\begin{align}
&g(T) \approx c_1+ \frac{3 \Omega_{w,0} T_0}{8 T^2} \sum _{n=0}^{\infty} \frac{1}{n!}\left[\frac{3 H_0 t_0 (1+w)}{1+\alpha}\right]^n \left[1+\dfrac{4 \alpha n}{1-3 \alpha}\left(\frac{T}{T_0}\right)^{\frac{1+\alpha}{2 \alpha}}\right],
\end{align}
By the use of the series expansions of exponentials, 
the solution above simplifies to a much simpler form
\begin{align}\label{eq_hubble_tgrescale_g_TT0_more}
&g(T) \approx c_1 + \frac{3 \Omega_{w,0} T_0}{8 T^2} \exp \left[\frac{3 H_0 t_0 (1+w)}{1+\alpha}\right] \left[1+\frac{3 H_0 t_0 (1+w)}{1+\alpha}\dfrac{4 \alpha}{1-3 \alpha}\left(\frac{T}{T_0}\right)^{\frac{1+\alpha}{2 \alpha}}\right].
\end{align}
Note that for a sufficiently small ratio $\left(\frac{T}{T_0}\right)^{\frac{1+\alpha}{2\alpha}}$, the last term in the square-bracket factor can be approximated to a constant. 

\end{itemize}

%

\section*{References}


\end{document}